\documentclass{jfm}
\usepackage{graphicx}
\usepackage{epstopdf, epsfig, amsmath, amssymb, bm}
\usepackage{booktabs}

\shorttitle{Shallow island wakes}
\shortauthor{P. M. Branson, M. Ghisalberti, G. N. Ivey and E. J. Hopfinger}
 
\title{Cylinder wakes in shallow oscillatory flow: the coastal island wake problem}
\author{Paul M.
Branson\aff{1}\corresp{\email{paul.branson@uwa.edu.au}}, M.
Ghisalberti\aff{1}, G. N. Ivey\aff{1} \and E. J. Hopfinger\aff{2}}
  
\affiliation{\aff{1}Oceans Graduate School,
University of Western Australia, Crawley, Australia
	\aff{2} LEGI, CNRS/UGA, Grenoble, France} 

\begin{document}

\maketitle

\begin{abstract}

	Topographic complexity on continental shelves is the catalyst that transforms the barotropic tide into the secondary and residual circulations that dominate vertical and cross-shelf mixing processes. Island wakes are one such example that are observed to significantly influence the transport and distribution of biological and physical scalars. Despite the importance of island wakes, to date, no sufficient, mechanistic description of
	the physical processes governing their development exists for the general case of unsteady tidal forcing. Controlled laboratory experiments are necessary for the understanding of this complex flow phenomena. 
	Three-dimensional velocity field measurements of cylinder wakes in shallow-water, oscillatory flow are conducted across a parameter space that is typical of tidal flow around shallow islands. Previous studies investigated the wake form dependance on $KC=U_0T/D$, where $KC$ is the Keulegan-Carpenter number, $D$ is the island diameter, $U_0$ the tidal velocity amplitude and $T$ the tidal period, and the stability parameter $S=c_fD/h$ where $h$ is the water depth and $c_f$ is the bottom boundary friction coefficient. In this study we demonstrate that when the influence of bottom friction is confined to a Stokes boundary layer the parameter $S=\delta^+/KC$ where $\delta^+=\delta/h$ and $\delta=2\pi\sqrt{2\nu/\omega}$ is the wavelength of the Stokes bottom boundary layer. Three classes of wake form are observed for decreasing wake stability: \emph{(1) Steady Bubble} for $S\gtrsim 0.1$; \emph{(2) Unsteady Bubble} for $0.06\lesssim S \lesssim 0.1$; and \emph{(3) Vortex Shedding} for $S\lesssim 0.06$. Transitions in wake form and wake stability are shown to depend on the magnitude and temporal evolution of the wake return flow. Scaling laws are developed to allow upscaling of the laboratory results to island wakes. Vertical and lateral transport depend on three parameters: 1) the flow aspect ratio $h/D$, where $h$ is the flow depth and $D$ the cylinder (island) diameter; 2) the amplitude of tidal motion relative to the island size, given by $KC$; and 3) the relative influence of bottom friction to the flow depth given by $\delta^+$.	A model of wake upwelling based on Ekman pumping from the bottom boundary layer demonstrates that upwelling in the near wake region of an island scales with $U_0(h/D)KC^{1/6}$ and is independent of the wake form. Finally, we demonstrate an intrinsic link between the dynamical eddy scales, predicted by the Ekman pumping model, and the island wake form and stability.

\end{abstract}

\section{Introduction}
Shallow-water tidal flow around islands or headlands produces recirculation
zones in their wake that is well documented by satellite images and aerial
photographs (e.g. \cite{Wolanski1984,Ingram1987}). These zones contribute
substantially to the vertical (upwelling) and horizontal material transport. Numerous studies have identified the importance of complex flow features 
such as eddies and fronts (shear zones) for aggregating sediment and 
plankton \citep{Wolanski1984,Ingram1987,Pattiaratchi1987}.
The upwelling of nutrient rich
water into the surface layer (that has higher light intensity) has important
ecological implications due to the increased primary productivity that results.
This in turn can effect the distribution of benthic and higher order pelagic organisms,
that graze along these fronts and wakes, taking advantage of lower-trophic level
food aggregations  \citep{Wolanski1988,Jenner2001,Johnston2007}. The secondary
circulation zones are complex, with a three dimensional structure characterised by
converging, diverging and curved flows \citep{White2008}. Observations of remotely sensed sea surface temperature data suggest that tidal mixing around islands in the Kimberley region of north western Australia is modulated by the spring-neap cycle and mixes sub-thermocline nutrient rich water to the surface in water depths of 60 - 80 m \citep{Creswell2000}. Whilst the potential ecological significance of shallow island wakes has been recognised for several decades, to date no quantitative description of the upwelling characteristics of shallow islands as a function of the flow parameters has been established. Laboratory experiments and numerical simulations that can determine the unsteady, 3D velocity fields play an essential role in this quest. We utilise a simple circular island cross section in this fundamental experimental study of the characteristics of shallow island wakes. 

Unbounded cylinder wakes in oscillatory flow have been studied extensively
(e.g. \cite{Sumer2006,Williamson1985}) and demonstrate the importance of the ratio of the flow excursion to cylinder size as described by the Keulegan-Carpenter number $KC=U_0T/D$, where $U_0$ is the amplitude of the sinusoidal velocity oscillation, $T$ the oscillation period and $D$ the cylinder diameter. The process of half-cycle interaction and pairing of vortices is fundamental to the observed wake patterns and, in turn, the fluid induced lift and in-line forces on the cylinder. 

In shallow water, experimental, field and theoretical studies of island wakes (assuming steady flow) have demonstrated the governing role of bottom friction on the wake structure expressed by the stability parameter $S=c_fD/h$ where $h$ is the flow depth and $c_f$ is a bed friction coefficient \citep{Chu1983,Chen1995,Lloyd1997,Negretti2006}. $S$ can be derived from the balance of the advective and frictional terms of the two-dimensional, depth-averaged momentum equation \citep{Pingree1980}. Laboratory studies of island wakes in turbulent, shallow flow have demonstrated that vortex shedding occurs for $S \lesssim 0.2$, an unsteady bubble wake for $0.2 \lesssim S \lesssim 0.5$ and, an attached, steady bubble wake for $S \gtrsim 0.5$. \cite{Wolanski1984} proposed a model of a wake recirculation region with an Ekman benthic boundary layer to describe the vorticity dynamics and characteristic length scale of the wake of Rattray Island in the Great Barrier Reef. That study proposed that the wake characteristics were governed by the island wake parameter $P=(Uh^2)/(\nu_TD)$ where $U$ is the steady flow velocity and $\nu_T$ is the turbulent vertical eddy viscosity. Vortex shedding is expected for $P >> 1$. Defining a quadratic bed friction coefficent as $c_f = (U/u_*)^2$ (where $u_*$ is the friction velocity) and assuming that in shallow, steady flow $\nu_T\sim u_*h$ \citep{Fischer1979} leads to $S\sim\sqrt{c_f}/P$ \citep{Chen1995}. For laboratory studies when the flow is laminar, the diffusion of momentum is controlled by the molecular viscosity $\nu$, the Reynolds number is $Re=UD/\nu$ and $P=Re(h/D)^2$ which is the controlling parameter for the cylinder wake in a Hele-Shaw cell \citep{Riegels1938}.

Laboratory experiments of \emph{oscillatory, shallow-water} island wakes are, however, rare. The wake forms and related horizontal velocity fields were determined from dye visualisation and PTV of surface floating particles by \cite{Lloyd2001}. Cylindrical and conical island geometries were invesitgated with wake form classification presented in the parameter space $S-KC$. Four modes of wake form were identified for both geometries with wake assymmetry increasing for decreasing $S$ and increasing $KC$. A 2D numerical simulation study of shallow-water headland wakes in tidal flow investigated a similar parameter space, a friction
Reynolds number equal to $1/S$, and the (dimensionless) wake length $KC/2\pi$ \citep{Signell1991}. That study also identified that the wake behaviour was predominantly governed through relative value of two length scales, a tidal excursion length $l_t=U_0T/2\pi$ and a frictional length $l_f=h/c_f$ \citep{Signell1991}. The island and headland wakes observed by \cite{Signell1991} and \cite{Lloyd2001} demonstrated the importance of bed friction on wake stability through the parameter $S$ and both studies conclude that the importance of transience in the wake development due to flow unsteadiness is quantified by $KC$. 

Whilst the wake stability controls the spatial distribution the wake eddies, the ecologically important process (that has yet to be characterised) is the vertical transport in the island wake. In shallow flow, it is frequently argued that the vertical velocity is constrained due to the small aspect ratio $h/D$. 
% The continuity equation in cartesian coordinates is:
% \begin{equation}
% 	\frac{\partial u}{\partial x} + \frac{\partial v}{\partial y} + \frac{\partial w}{\partial z} = 0
% 	\label{eqn:com}
% \end{equation}
% where $u$, $v$ and $w$ is the velocity in the $x$, $y$ and $z$ directions, respectively. Taking the island diameter $D$ as a characteristic horizontal length scale and the flow depth $h$ as a characteristic vertical length scale, Equation \ref{eqn:com} implies:
% \begin{equation}
% 	O(w) \sim \frac{h}{D}O(u)
% 	\label{eqn:com2}
% \end{equation}
% where $O(u)$ and $O(w)$ are the order of magnitude of the horizontal and vertical velocity respectively. 
Despite this constraint, experimental studies on shallow vortices \citep{Akkermans2008a}, and more recently, shallow island wakes \citep{Branson2018a} have observed larger than expected vertical velocities and complex interactions between primary and secondary vortices. The near-bed stress at the base of a shallow vortex forms an Ekman boundary layer that is able to actively suck (and pump) fluid into (and out of) the benthic boundary layer \citep{Greenspan1968}. Numerical and analytical studies of decaying shallow vortices have demonstrated that the upwelling regime is controlled by the relative bottom boundary layer thickness within the vortex \citep{DuranMatute2012}. However, an analytical solution for the scaling of upwelling was only possible in the viscous regime where vertical diffusion of momementum dominated. In contrast to isolated shallow vortices, in the island wake there is an interaction between the wake eddy structure and the scales of the external tidal flow \citep{Branson2018a}. A mechanistic description of the interaction between the external tidal flow field and the circulation in the island wake is needed to improve understanding of the topographically induced mixing.

The present paper utilises a novel experimental approach with three main objectives: (i) to determine consistent wake 
classification criteria for shallow, oscillatory flow around cylinders; (ii) to establish the 3D near wake flow field; and (iii) to establish scaling laws for lateral and upwelling velocities and the onset of vortex shedding.  

\section{Wake classification parameters}\label{sec:wakeClass}

We demonstrate here that there exists an inverse dependence between $S$ and $KC$ that confounds the relative influence of water depth and transient bed
friction (due to the Stokes boundary layer development). Furthermore, the relative length scales identified in \cite{Signell1991} establish an additional parameter that directly links $KC$ and $S$ and expresses the role of vertical length scales on the wake characteristics. 

The well known Stokes boundary layer solution is \citep{Batchelor1967}:
\begin{equation}
u(z,t)=U_0[\cos(\omega t)-e^{-kz}\cos(\omega t-kz)]
\label{eqn:stokes}
\end{equation}
where $U_0$ is the velocity amplitude outside of the boundary layer, $\omega$ is the radial frequency, $k=\sqrt{\omega/2\nu}$ is the wave number and,
$\delta=2\pi\sqrt{2\nu/\omega}$ is the wavelength, also referred to as the Stokes
boundary layer thickness. Similarly, the boundary layer thickness defined by
$\partial u(z,t)/\partial z=0$, is
$\delta_{BL}=\frac{3}{4}\pi\sqrt{2\nu/\omega}$. The shear stress is $\tau(z,t)=\mu\frac{\partial u(z,t)}{\partial z}$ where, for laminar oscillatory flow over a smooth boundary (which is the
case in the present experiments), the bed shear stress (at $z=0$) is $\tau(t)=\mu
U_0\sqrt{\omega/2\nu}(\cos\omega t-\sin\omega t)$. The maximum bed friction coefficient is
\begin{equation}
c_f=\frac{\tau_0}{\frac{1}{2}\rho U_0^2}=\frac{2\nu/U_0}{\sqrt{2\nu/\omega}}
\label{eqn:cf}
\end{equation}
where $\tau_0$ is the peak in bed shear stress. This bed friction coefficient can be utilised in the calculation of $S$.
Substituting the expression for $c_f$ into that for $S$ yields:
\begin{equation}
    S=\frac{\delta/h}{U_0T/D}
    %=\frac{1}{KC(h/\delta)}
    \label{eqn:1onS}
\end{equation}
demonstrating that $S$ is dependent on both the relative wake length $KC=U_0T/D$ and the ratio of boundary layer wavelength to flow depth $\delta^+=\delta/h$. 

It is of interest to point out that $S$ can also be expressed in terms of a Reynolds number ($Re=U_0D/\nu$):
\begin{equation}
S=\frac{4\pi}{\delta^+Re(h/D)^2}
\end{equation}
which demonstrates that $\delta^+$ is the non-dimensional parameter that links $S$, $KC$ and $Re(h/D)^2$ (i.e. $P$ with $\nu_T=\nu$) for \emph{oscillatory, shallow} wake flow.  Further, in terms of the parameters observed to predominantly govern headland wake behavior ($l_f$ and $l_t$), if $c_f$ is determined using Equation \ref{eqn:cf}, $\delta^+$ is equivalent to
$l_f/l_t$, consistent with the conclusions of \cite{Signell1991}. The parameter $\delta^+$ describes the relative scale of influence of bottom friction compared to the flow depth. When $\delta^+>>1$ the influence of friction is large and when $\delta^+<<1$ the influence of friction is small. 

The two parameters $KC$ and $\delta^+$
that determine the stability parameter $S$ in oscillatory island wake flow can be independently varied
and allows us to examine the influence of relative wake length and bed friction on the wake of a shallow island in a sinusoidal tidal flow.

\section{Methods}\label{sec:expSetup}

\subsection{Experimental conditions}

A total of 38 experiments were completed in this study (table \ref{tab:runs}). Circular cylinders
of diameter 10 cm and 15 cm were utilised as an analogue for an island, with the flow depth ($h$)
varied from 2.3 cm to 5.3 cm, velocity amplitude ($U_0$) from 0.4 cm~s$^-1$ to 6.2 cm~s$^-1$ and oscillation period ($T$) from 60 s to 177 s. This covered a wide range in the controlling parameters: $4.5 < KC < 48.0$, $0.5 < \delta^+ < 2.0$ and $0.02 < S < 0.36$ including several experiments with repeated $KC$ and $\delta^+$ but with differing tidal period ($T$) and cylinder diameter ($D$).

\begin{table}
	\begin{center}
		\def~{\hphantom{0}} 
		% \makebox[1.0\textwidth][c]{
		\begin{tabular}{llllllllllll}
			% \toprule 
			Run &  $U_0$ & $T$ & $D$  & $h$  & $\delta$  & $\delta^+$ & $KC$ & $S$ & $Re$ & Wake type & Fig. \ref{fig:streamLinePlanView} ref. \\
			 &  [cm.s$^{-1}$] & [s] & [cm] & [cm] & [cm] &  &  &  &  & & \\[3pt]
			% \midrule
			1 & 4.4 & 60.0 & 10 & 5.3 & 2.7 & 0.52 & 26.5 & 0.02 & 4417 & VS & t \\
			2 & 2.4 & 60.0 & 10 & 5.3 & 2.7 & 0.52 & 14.3 & 0.04 & 2391 & VS & r \\
			3 & 3.5 & 60.0 & 10 & 5.3 & 2.7 & 0.52 & 21.1 & 0.02 & 3521 & VS & s \\
			6 & 1.7 & 120.0 & 10 & 5.3 & 3.9 & 0.73 & 20.2 & 0.04 & 1682 & VS & q \\
			7 & 1.0 & 120.0 & 10 & 5.3 & 3.9 & 0.73 & 11.4 & 0.06 & 953 & VS & p \\
			8 & 0.4 & 120.0 & 10 & 5.3 & 3.9 & 0.73 & 5.0 & 0.15 & 414 & SB & o \\
			9 & 2.1 & 121.6 & 10 & 4.3 & 3.9 & 0.91 & 25.8 & 0.04 & 2118 & VS & n \\
			10 & 1.2 & 121.6 & 10 & 4.3 & 3.9 & 0.91 & 14.4 & 0.06 & 1186 & UB & m \\
			11 & 0.8 & 121.6 & 10 & 4.3 & 3.9 & 0.91 & 9.7 & 0.09 & 797 & UB & l \\
			12 & 0.4 & 121.6 & 10 & 4.3 & 3.9 & 0.91 & 4.8 & 0.19 & 394 & SB & k \\
			13 & 2.5 & 121.6 & 10 & 3.3 & 3.9 & 1.18 & 29.8 & 0.04 & 2454 & VS &  \\
			14 & 1.5 & 121.6 & 10 & 3.3 & 3.9 & 1.18 & 17.7 & 0.07 & 1455 & UB & i \\
			15 & 0.9 & 121.6 & 10 & 3.3 & 3.9 & 1.18 & 11.5 & 0.10 & 946 & UB & g \\
			16 & 0.4 & 121.6 & 10 & 3.3 & 3.9 & 1.18 & 5.2 & 0.23 & 425 & SB & f \\
			17 & 1.3 & 176.9 & 10 & 3.3 & 4.7 & 1.43 & 22.2 & 0.06 & 1253 & UB & d \\
			18 & 1.0 & 176.9 & 10 & 3.3 & 4.7 & 1.43 & 17.1 & 0.08 & 969 & UB & h \\
			19 & 0.7 & 176.9 & 10 & 3.3 & 4.7 & 1.43 & 12.2 & 0.12 & 688 & SB & b \\
			20 & 0.3 & 176.9 & 10 & 3.3 & 4.7 & 1.43 & 5.9 & 0.24 & 332 & SB &  \\
			21 & 2.8 & 85.9 & 10 & 2.3 & 3.3 & 1.43 & 23.7 & 0.06 & 2764 & VS & j \\
			22 & 2.2 & 85.9 & 10 & 2.3 & 3.3 & 1.43 & 18.6 & 0.08 & 2165 & UB &  \\
			23 & 1.2 & 85.9 & 10 & 2.3 & 3.3 & 1.43 & 10.4 & 0.14 & 1208 & SB &  \\
			24 & 0.5 & 85.9 & 10 & 2.3 & 3.3 & 1.43 & 4.5 & 0.32 & 524 & SB &  \\
			25 & 2.0 & 116.9 & 10 & 2.3 & 3.8 & 1.67 & 23.6 & 0.07 & 2019 & UB & e \\
			26 & 1.5 & 116.9 & 10 & 2.3 & 3.8 & 1.67 & 17.6 & 0.09 & 1504 & SB & c \\
			27 & 1.0 & 116.9 & 10 & 2.3 & 3.8 & 1.67 & 11.9 & 0.14 & 1019 & SB &  \\
			28 & 0.5 & 116.9 & 10 & 2.3 & 3.8 & 1.67 & 5.4 & 0.31 & 459 & SB & a \\
			29 & 3.7 & 116.9 & 15 & 2.3 & 3.8 & 1.67 & 28.8 & 0.06 & 5544 & UB &  \\
			30 & 3.0 & 116.9 & 15 & 2.3 & 3.8 & 1.67 & 23.0 & 0.07 & 4426 & UB &  \\
			31 & 1.6 & 116.9 & 15 & 2.3 & 3.8 & 1.67 & 12.1 & 0.14 & 2335 & SB &  \\
			32 & 0.7 & 116.9 & 15 & 2.3 & 3.8 & 1.67 & 5.5 & 0.30 & 1058 & SB &  \\
			33 & 2.1 & 168.4 & 15 & 2.3 & 4.6 & 2.00 & 23.5 & 0.09 & 3142 & UB &  \\
			34 & 1.6 & 168.4 & 15 & 2.3 & 4.6 & 2.00 & 18.5 & 0.11 & 2470 & SB &  \\
			35 & 1.1 & 168.4 & 15 & 2.3 & 4.6 & 2.00 & 12.1 & 0.17 & 1619 & SB &  \\
			36 & 0.5 & 168.4 & 15 & 2.3 & 4.6 & 2.00 & 5.5 & 0.36 & 735 & SB &  \\
			37 & 4.1 & 168.4 & 15 & 2.3 & 4.6 & 2.00 & 45.8 & 0.04 & 6116 & VS &  \\
			38 & 3.0 & 168.4 & 15 & 2.3 & 4.6 & 2.00 & 34.0 & 0.06 & 4550 & UB &  \\
			39 & 6.2 & 116.9 & 15 & 2.3 & 3.8 & 1.67 & 48.2 & 0.03 & 9283 & VS &  \\
			40 & 5.0 & 116.9 & 15 & 2.3 & 3.8 & 1.67 & 39.1 & 0.04 & 7530 & VS &  \\
			% \bottomrule
		\end{tabular}
		% }
		\caption{Conditions of the experimental runs, $\delta=2\sqrt{\pi \nu T}$, $KC=U_0T/D$,
		$\delta^+=2\sqrt{\pi\nu T}/h$, $S=\delta^+/KC$ and
		$Re=U_0D/\nu$. SB = Steady Bubble, UB = Unsteady Bubble and VS = Vortex Shedding. The final column provides a cross-reference for the subplots shown in figure \ref{fig:streamLinePlanView}.
		}
		\label{tab:runs}
	\end{center}
\end{table}

\subsection{Shallow tidal flow flume}

Experiments were conducted in a recirculating glass flume with a working
section 1.85 m wide, 6.00 m long and up to 0.35 m deep. The working section was
elevated to allow optical access from below. The flume had a computer controlled variable frequency reversible pump that generated a
barotropic reciprocating tidal flow of specified period and amplitude. The pump
and drive system was able to generate a near-sinusoidal velocity
signal. An array of polycarbonate flow straighteners and sponges were installed at each end
of the flume to produce a near-uniform transverse velocity profile. 

\begin{figure}
	\centerline{\includegraphics[width=1.0\textwidth]{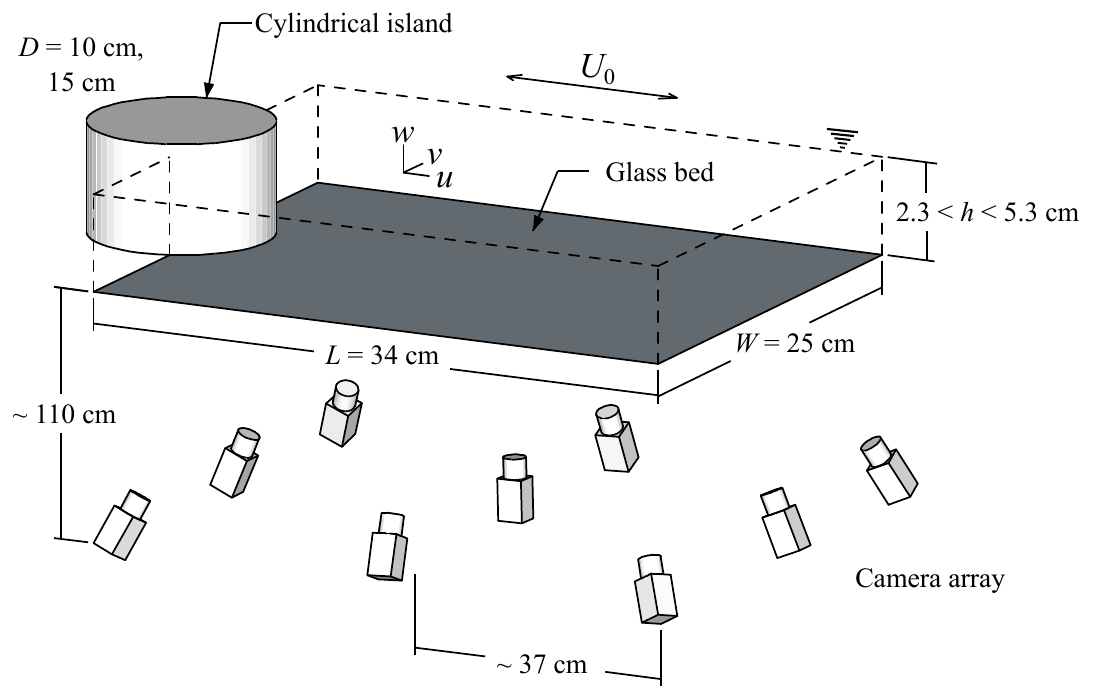}}% Images in 100% size
	\caption{Schematic diagram of the experimental setup for the 10 cm cylinder. The measurement volume (of length $L$ and width $W$) is indicated by the dashed lines, with the island located at its edge. Oscillating sinusoidal flow was generated with a velocity amplitude of $U_0$. The deepest and shallowest experimental depth conditions were 5.3 cm and 2.3 cm respectively. A synthetic aperture camera array imaged neautrally-buoyant Pliolite AC80 particles through the clear glass bed, with velocity estimates subsequently obtained using 3DPIV. Note that the camera array is not shown to scale.}
	\label{fig:ExperimentalSetup}
\end{figure}

A schematic of the experimental setup is shown in figure \ref{fig:ExperimentalSetup}. The working fluid was salt water, buoyancy matched and seeded with
300-450 $\mu$m
size particles of Pliolite AC80. The fluid was illuminated from the side with an
array of high intensity light emitting diodes. The flow was imaged with an
array of 9 Basler ACE acA1600-20gm 2.0 mega-pixel monochrome GigE cameras with a
maximum frame-rate of 20 fps. The camera array was positioned below the flume
at a distance of approximately 110 cm. The cameras were arranged in a 3 x 3 grid
layout with approximately 37 cm spacing between each camera. The cameras were
aligned on a 34 x 25 x 6 cm{$^3$} measurement volume. 

\subsection{Synthetic apeture particle imaging velocimetry}

The seeded flow was imaged from multiple points of view using the camera array and the three-dimensional volume of particles was reconstructed from the two dimensional camera images for three-dimensional particle imaging velocimetry
(3DPIV). In this study we applied the synthetic aperture imaging technique for volume reconstruction. If the geometry of the imaging system is well known, it is possible to
re-parameterise the scene \citep{Isaksen2000} into a number of discrete focal
planes located arbitrarily in the imaged volume using the Map-shift-average
algorithm \citep{Vaish2005}. \cite{Belden2010} evaluated the synthetic aperture
imaging technique for use in 3DPIV and found that the reconstruction quality increased with the number of cameras (up to approximately 12).
The multi-camera self-calibration technique of \cite{Belden2013} was utilised to establish the geometry of the camera array,
with a mean reprojection error of 0.11 px across the 9 cameras in the array. The resulting horizontal and
vertical size of each reconstucted voxel was 0.22 mm and 0.5 mm respectively, with a total
volume size of 193.6 million voxels for the deepest flow condition ($h=5.3$
cm). 

After reconstruction of consecutive volumes an iterative, multi-grid, image deforming 3DPIV algorithm was applied with
ensemble averaging in the correlation space to overcome non-uniformity in
seeding density \citep{Raffel2007} (due to the converging and diverging flows).
Sub-pixel displacement estimates were obtained using a three point Gaussian fit
to the correlation peak in three dimensions. The universal outlier detection
algorithm of \cite{Westerweel2005} was applied after each pass to remove
spurious vectors. Finally, a three-dimensional Gaussian filter ($\sigma=0.5$) was applied to reduce noise. 

The iterative analysis was applied for a total of 8 passes, with the first three
passes progressively refining the interrogation volume from 256 x 256 x 48
voxels through to 96 x 96 x 12 voxels with 75\% overlap. This resulted in a
three-dimensional vector field of 64 x 48 x 32 (98304) vectors sampled at 1 Hz with physical spacings of 5.3 mm horizontally and 1.5
mm vertically. Further details of the technique and validation of the 3D
velocity measurements against a co-located ADV are provided in \cite{Branson2018b}.

\subsection{Measurement uncertainty}

Particle imaging velocimetry measurement uncertainty has previously been investigated \emph{a priori} using theoretical modelling \citep{Westerweel1997} and Monte Carlo simulation of the measurement chain (e.g. \cite{Keane1992} and others). These analyses have helped optimise the various aspects of the measurement chain, with most studies estimating error at approximately 0.1 px \citep{Raffel2007}. Due to continuity, the divergence should be zero everywhere (i.e. $\nabla\bullet\vec{u} = 0$) in an incompressible flow. Thus, the RMS of the divergence (typically normalised by observed velocity gradients or vorticity) can be evaluated as a measure of experimental error (e.g. \cite{Scarano2009,Atkinson2011}). This approach allows for the measurement error to be assessed \emph{a posteriori} and aggregates the systematic and random errors associated with the experimental apparatus and PIV algorithm. Previous studies have applied uncertainty analysis \citep{Moffat1988} and established that error in the velocity gradient can be expressed in terms of the random velocity error \citep{Atkinson2011,Earl2013}:
\begin{equation}
  \epsilon(\nabla\bullet\vec{u})=\sqrt{\frac{3}{2\Delta^2}}\epsilon(u)
  \label{DivError}
\end{equation}
where $\Delta$ is the characteristic length scale of the 3D vector grid (in pixels). Whilst Equation \ref{DivError} provides an estimate of the absolute error relative to the PIV measurement resolution, we also assess the RMS divergence relative to the magnitude of the measured vorticity; the latter is taken as the $99^{th}$ percentile of the vorticity vector norm ($\|\boldsymbol{\omega}\|_{99}$ where $\boldsymbol{\omega}=\nabla\times\vec{u}$) for each experimental condition. The inter-frame time for the PIV analysis was adjusted for each experimental condition to minimise the random velocity error and relative divergence error. The velocity error ($\epsilon(\nabla\bullet\vec{u})$) varied from 1 - 4\% and vorticity error ($\nabla\bullet\vec{u}/\|\boldsymbol{\omega}\|_{99}$) from 8 - 17\%, comparable to other 3DPIV studies of cylinder wakes \citep{Scarano2009}.

\subsection{Data analysis}
\begin{figure}
	\centerline{\includegraphics[width=1.0\textwidth]{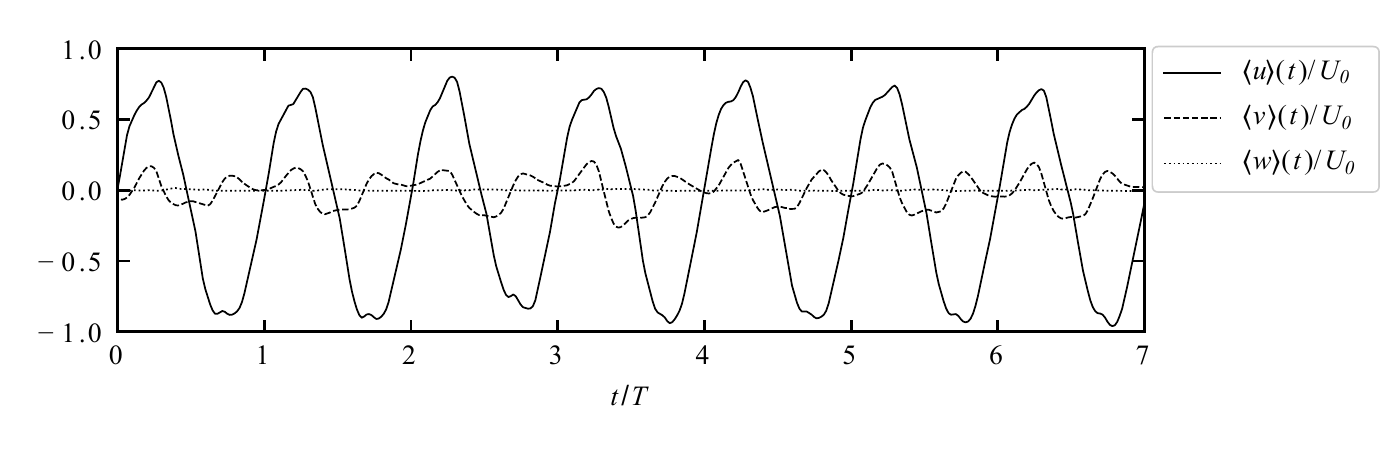}}% Images in 100% size
	\caption{Example of time variation of the volume averaged velocities $\langle u\rangle(t)$,
	$\langle v\rangle(t)$ and $\langle w\rangle(t)$ for Run 1. Peak flow past the island occurs at $\phi =
	0.25$ and $0.75$ and the wake is generated in the measurement volume for
	$0.0 < \phi < 0.5$. The amplitude of $\langle u\rangle(t)$ does not reach $U_0$ as $U_0$ is defined at a height of $z = \delta_{BL}$, whilst $\langle u \rangle(t)$ is the volume average. Note the deficit in amplitude of $\langle u\rangle(t)$ during the wake generation half-cycle.}
	\label{fig:Run01FieldAvg}
\end{figure}

In the presentation of the results, spatial averages are denoted by $\langle ...\rangle$, long-term averages by an over-bar (e.g. $\overline{u}$) and phase averages by a tilde (e.g. $\tilde{u}$) (for a particular tidal phase). Subscripts are utilised to indicate if a statistic other than the mean is calculated. For example, the time series of root mean square velocity is calculated as
\begin{equation}
\langle u \rangle_{rms} (t) =\sqrt{\frac{1}{N}\sum^{x,y,z}u^2}
\end{equation}
where $N$ is the number of velocity samples within the measurement volume. 
%In an unbounded oscillatory flow, a boundary layer will develop with thickness $\delta_{BL}=\frac{3}{4}\pi\sqrt{2\nu/\omega}$ \citep{Batchelor1967}. 
The velocity, $u_{BL}$, at the $\delta_{BL}$ boundary layer height is calculated as the spatially-averaged horizontal velocity $\langle u\rangle$ at a height $z=\delta_{BL}$ for $x/D > 2.5$. The amplitude of the sinusoidal velocity ($U_0$) is calculated from the $u_{BL}$ timeseries upstream of the cylinder. Due to the presence of the Stokes boundary layer there is a phase shift in the horizontal velocity with distance from the bed. To phase align the results, the time $t=0$ is defined at the first flow zero-crossing of $u_{BL}$. Thus, the phase of each experiment is defined as $\phi=t/T ~\textrm{mod}~1$. The wake `growth' period is $0.0 < \phi < 0.5$ and the wake `decay' period $0.5 < \phi < 1.0$. An example of the volume averaged velocity for Run 1 is presented in figure \ref{fig:Run01FieldAvg}. For this vortex shedding wake, perturbations in $\langle u \rangle$ and $\langle v \rangle$ are observed due the presence of wake eddies and wake oscillation. The value of spatially-averaged $\langle w \rangle$ is approximately zero, as would be expected by conservation of mass in a shallow flow.

In this paper, we develop scaling relationships for the root mean square lateral $\langle \overline{v}\rangle_{rms}$ and vertical $\langle \overline{w}\rangle_{rms}$ velocities. These scaling relationships are established based on power laws. In log space, a power law can be represented as a general linear model of the form:
\begin{equation}
	\bm{Y}_i = \beta _0 + \beta _1 \boldsymbol{X} _{i1} + \beta  _2 \boldsymbol{X}_{i2} + ... + \beta _p \boldsymbol{X }_{ip} + \boldsymbol{\epsilon} _i
\end{equation}
where $\boldsymbol{Y}$ is a series of $i$ observations, $\boldsymbol{X}$ is a matrix of $p$ predictors (i.e. $KC$, $\delta^+$ etc.), $\beta$ are scaling coefficients to be estimated and $\boldsymbol{\epsilon} _i$ is a series of errors assumed to follow a log-normal distribution. The scaling coefficients and intercept ($\beta _0$) can be estimated using ordinary linear regression with \emph{p}-values calculated using the $t$-test and 95\% confidence intervals of parameter estimates based on the Student's $t$-distribution. However, ordinary linear regression only provides the single maximum likelihood estimate of the scaling coefficients. A Hamiltonian Monte Carlo (HMC) algorithm is applied to estimate the posterior distribution of the scaling coefficents. HMC is a Markov-Chain Monte Carlo (MCMC) algorithm that is particularly suited to problems where correlation between predictors can be problematic for many MCMC methods. The No-U-Turn sampler \citep{Hoffman2014} as implemented in the PyMC3 library \citep{Salvatier2016} is utilised to sample the posterior distributions of the scaling coefficents. For all MCMC results presented in this study, four chains of 4000 samples with an initial tuning of 1000 samples were executed. Tuning samples were discarded and the potential for biased parameter estimates checked by examining the trace auto-correlation and trace divergence. The results of this analysis are presented in the Supplementary Material.

\section{Results}\label{sec:results}

\subsection{Velocity variations with $h/D$, $\delta^+$ and $KC$}

\begin{figure}
	\centerline{\includegraphics[width=1.0\textwidth]{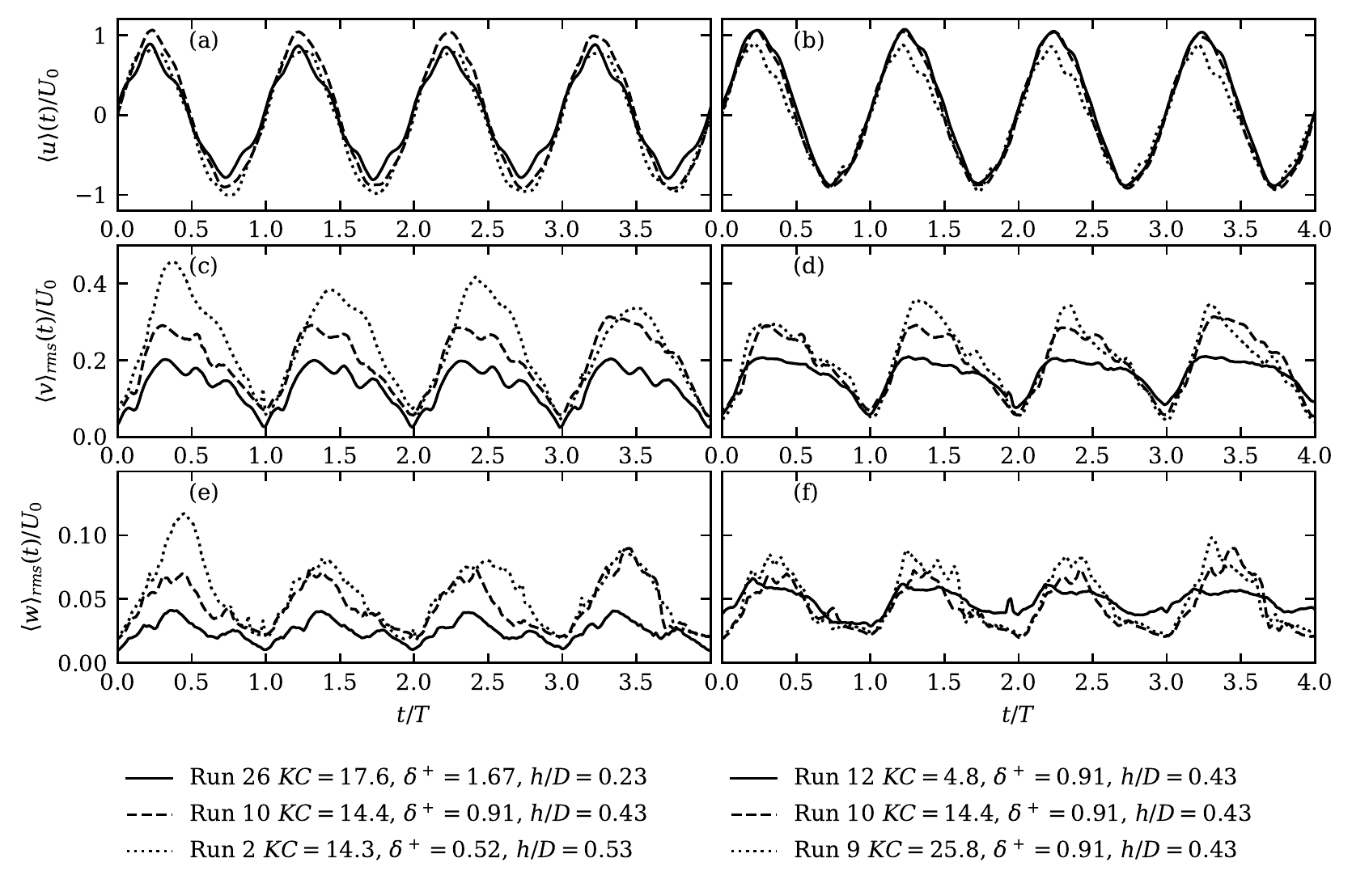}}% Images in 100% size
	\caption{Example variation in ${\langle u \rangle (t)}/U_0$, $\langle v \rangle _{rms}(t)/U_0$ and $\langle w \rangle _{rms}(t)/U_0$ with
	$KC$ and $\delta^+$. The left column of figures shows approximately constant $KC\approx 15$ and variable $\delta^+$ and the right column of figures shows constant $\delta^+ = 0.91$ and variable $KC$. There is an increase in magnitude and cycle-to-cycle variability in $\langle v \rangle _{rms}$ with decreasing $\delta^+$. The dependence of $\langle v \rangle _{rms}$ on $KC$ is weak compared to the dependence of $\langle v \rangle _{rms}$ on $\delta^+$. The vertical velocity $\langle w \rangle _{rms}$ increases with increasing $h/D$ with a weak dependance on $KC$.}
	\label{fig:FieldAvg_Comp}
\end{figure}

Field-averaged measures of $u$, $v$ and $w$ provide a simple view of the overall wake dynamics, time variability and sensitivity to the controlling parameters $h/D$, $KC$ and $\delta^+$. The variation of $\langle u \rangle(t)/U_0$, $\langle v \rangle _{rms}/U_0$ and $\langle w \rangle _{rms}/U_0$ with
$KC$ and $\delta^+$ is shown in figure \ref{fig:FieldAvg_Comp}. The left column shows $KC\approx15$ and the right column $\delta^+=0.91$. For these flow conditions, changes in $\delta^+$ and $h/D$ are negatively correlated. However, this is not the case in the larger dataset of 38 flow conditions. This subset of 5 flow conditions will be utilised to make some general observations that will inform the development of the scaling arguments.

As expected, the depth-averaged parameter $\langle u \rangle$ decreases with increasing $\delta^+$ due to the increasing boundary layer thickness (figure \ref{fig:FieldAvg_Comp}a). The velocity deficit during the wake growth period increases with increasing $KC$ (figure \ref{fig:FieldAvg_Comp}b). Decreasing $\delta^+$ results in an increase in the magnitude and cycle to cycle
variability of the normalised value of $\langle v \rangle _{rms}$ (figure \ref{fig:FieldAvg_Comp}c). By contrast, $KC$ has a lesser influence on $\langle v \rangle _{rms}$, with a non-linear increase in $\langle v \rangle _{rms}$ from $KC=4.8$ to $KC=14.4$, but minimal increase from $KC=14.4$ to $KC=25.8$ (figure \ref{fig:FieldAvg_Comp}d). The magnitude of $\langle w \rangle _{rms}$ increases with increasing $h/D$ and decreasing $\delta^+$ (figure \ref{fig:FieldAvg_Comp}e), with a weak dependence on $KC$ (figure \ref{fig:FieldAvg_Comp}f). For the $KC=4.8$ example, $\langle w \rangle _{rms}$ is smaller during the peak generation of the wake, but sustained throughout the flow reversal and remainder of the tidal cycle. As observed in \cite{Branson2018a}, this is due to vertical velocity associated with long, $x$-aligned vortices that persist through the tidal cycle. These observations qualitatively support the hypothesis that the lateral and vertical velocity are governed by $h/D$, $KC$ and $\delta^+$.

\subsection{Wake classification}
In steady flow around an island, flow separation
leads to the formation of a wake `bubble', a region of low pressure in the lee
of the island which results in the generation of counter-rotating vortices. When
the counter-rotating vortices remain attached to the island, the wake is
considered stable. In oscillating flow, wake stability has been considered
less meaningful than in steady flow due to the free stream being unsteady which can
result in the convection of vorticity back past the island \citep{Lloyd2001}.
The convection of vorticity and half-cycle interaction lead to a wide range of
wake regimes and vortex synchronisation patterns in unbounded oscillatory flow past a cylinder \citep{Williamson1988}.

The dominance of the bottom boundary layer in shallow oscillatory flow ($h\ll D$)
fundamentally alters the wake characteristics due to the generation of
vorticity in the bottom boundary layer and momentum transfer from the wake
to the bed. Previous investigations in shallow, oscillatory island wakes have utilised four
categories: \emph{(i) symmetric without vortex pairing; (ii) symmetric with
pairing; (iii) sinuous with pairing; and (iv) vortex shedding} \citep{Lloyd2001}.
In this study the wake classification of \cite{Branson2018a} has been utilised as it more clearly articulates the similarity to the stability criteria of shallow, steady wakes presented in \cite{Chen1995,Chen1997}. The four classes of \cite{Branson2018a} were: \emph{(i) symmetric; (ii) asymmetric; (iii) unsteady bubble;} and \emph{(iv) vortex shedding}. 

\begin{figure}
	\centerline{\includegraphics[width=0.95\textwidth]{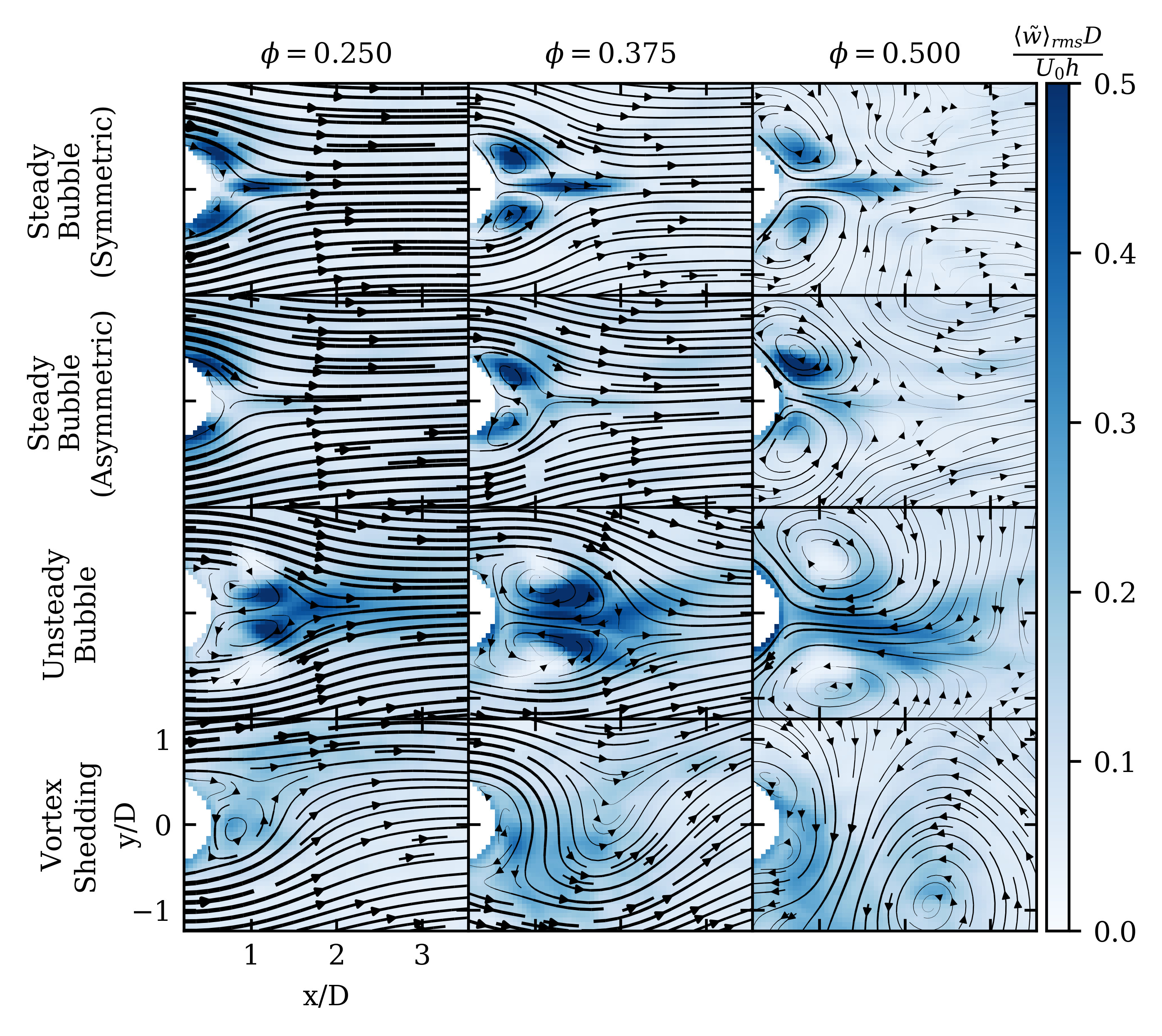}}% Images in 100% size
		\caption{Evolution of oscillatory shallow island wake for $0.25 < \phi < 0.5$. Examples of a symmetric (Run 24), and asymmetric (Run 12) \emph{steady bubble (SB)} wake. An \emph{unsteady bubble (UB)} wake (Run 14), and, an unstable \emph{vortex shedding (VS)} wake (Run 3). Streamlines show the depth- and phase-averaged velocity in the reference
		frame of the island. The streamline width indicates the relative magnitude of the horizontal velocity (normalised by $U_0$). Shaded areas of large vertical advection shown by the normalised RMS vertical velocity $\frac{\langle \tilde{w} \rangle_{rms} D}{U_0h}$.}
		\label{fig:wakeClassification}  
	\end{figure}

Examples of the streamline topology of the different wake classes are demonstrated through the
depth- and phase-averaged 3D PIV data presented in figure \ref{fig:wakeClassification}. Wake forms were classified based on visual observations during the experiments and through animation of instantaneous, depth-averaged velocity fields. We group the symmetric and asymmetric wakes into a single class: steady bubble. The asymmetric wake is present only for small $KC$ and low $\delta^+$ and is distinguished from the unsteady bubble wake by the lack of strong downwelling secondary vortices. Whilst the asymmetric wake has one wake eddy with a larger circulation than the other, it repeats consistently from cycle-to-cycle with little unsteadiness in the wake bubble within a half-cycle. Deflection of the secondary vortices (indicated by regions of enhanced vertical velocity) away from the wake centreline ($y/D=0$) marked the transition from the steady bubble wake to the unsteady bubble wake. Finally, the transition to vortex shedding was indicated by the detachment of a primary vortex during the `growth' period. Thus, the wake classes utilised here are \emph{(1) Steady Bubble (SB), (2) Unsteady Bubble (UB),} and \emph{(3) Vortex Shedding (VS)}. Visually, the laminar wakes of this study are markedly different from the (equivalently-named) turbulent wakes of \cite{Chen1995}. However, as will be demonstrated later, the wake classes are analogous from the perspective of the strength of the wake return flow and global stability of the wake.

\begin{figure}
	\centerline{\includegraphics[width=1.0\textwidth]{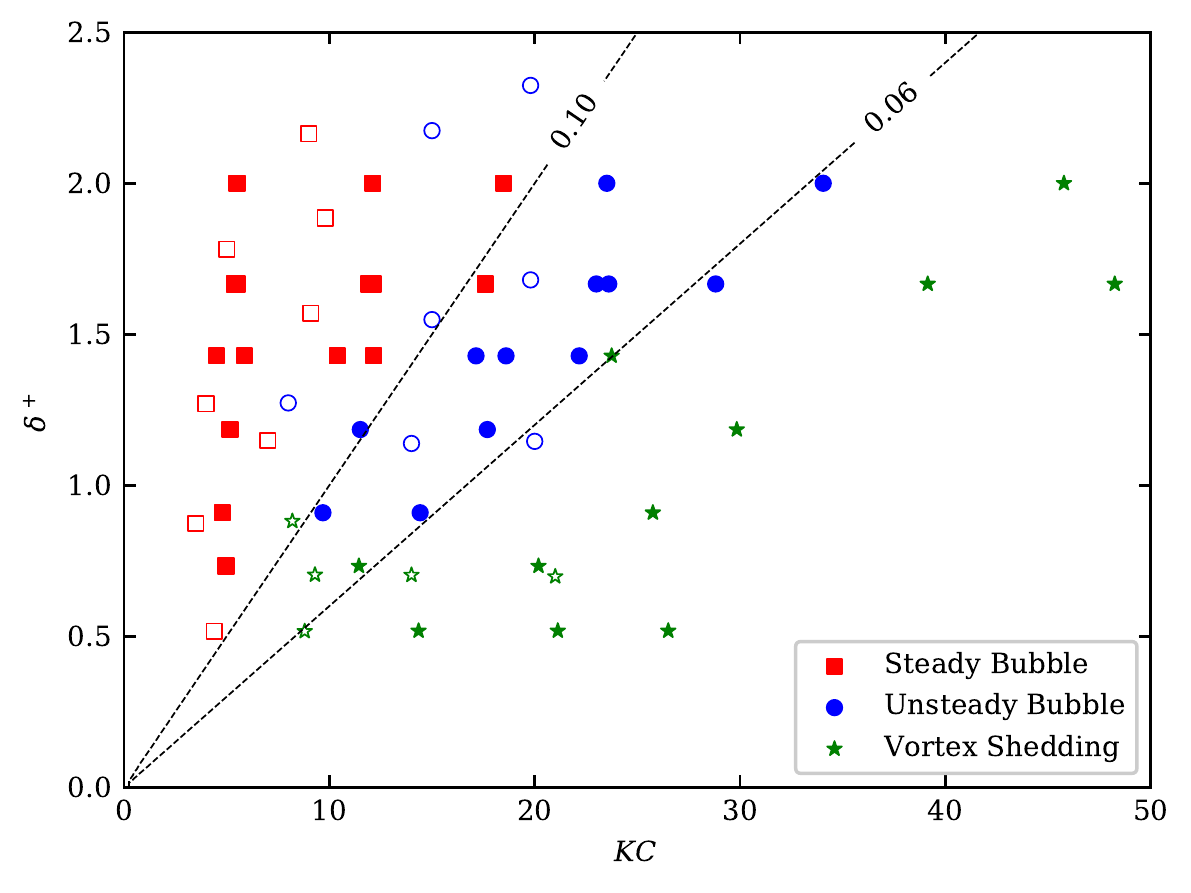}}% Images in 100% size
	\caption{Presentation of the experimental conditions in $KC$-$\delta^+$
	space with labelled lines of constant $S$ shown. Filled markers refer to conditions in
	the present study and unfilled markers are those of \cite{Lloyd2001}. 
	% Numbers correspond to Run identifier in table \ref{tab:runs}. 
	Marker type indicates the wake
	form. 
	% :
	% symmetric ($\bullet$), asymmetric ($\blacktriangleleft$) and vortex shedding ($\blacksquare$). The transition condition to vortex shedding, discussed in
	% section
	% \ref{sec:trans}, is indicated by a black dashed line.
	}
	\label{fig:ParamSpace}
\end{figure}

In this	study, vortex shedding was observed for $S\lesssim0.06$, the unsteady bubble wake for $0.06 \lesssim S\lesssim0.1$ and the steady bubble wake for $S\gtrapprox 0.1$. The transition from one wake class to the next was observed to be gradual (figure \ref{fig:ParamSpace}). Experiments close to the transition to vortex shedding were observed to have
cycle-to-cycle variability. That is, asymmetry was observed to grow over a
number of cycles which may culminate in a vortex being shed only every few
cycles. There is a complex inter-dependence
between $KC$ and $\delta^+$ that governs the transition between wake classes. 
The pairing process documented in \cite{Lloyd2001} (where a vortex from the previous half-cycle is advected back past the cylinder and pairs with a newly forming vortex on the next cycle) is observed in the
present study for the SB wake (for $KC > 7$) and UB wake. However, due to the reduced field of
view in these experiments, the vortex pairs advect beyond $x/D = 3.5$ for large $KC$. Figure
\ref{fig:ParamSpace} shows the wake classification, in conjunction with the experimental values of $S$, in
$KC$-$\delta^+$ space. Here, the results of \cite{Lloyd2001} have been reclassified by grouping the \emph{symmetric without vortex pairing} and \emph{symmetric with
pairing} wakes as \emph{steady bubble}, and the \emph{sinuous with
pairing} wake as \emph{unsteady bubble} (noting also that the stability parameter of \cite{Lloyd2001} has been recalculated here by dividing by $\sqrt{2}$ to reconcile differing definitions of $c_f$). There is reasonable agreement between the results of \cite{Lloyd2001} and this study.

\begin{figure}
\centerline{\includegraphics[width=1.0\textwidth]{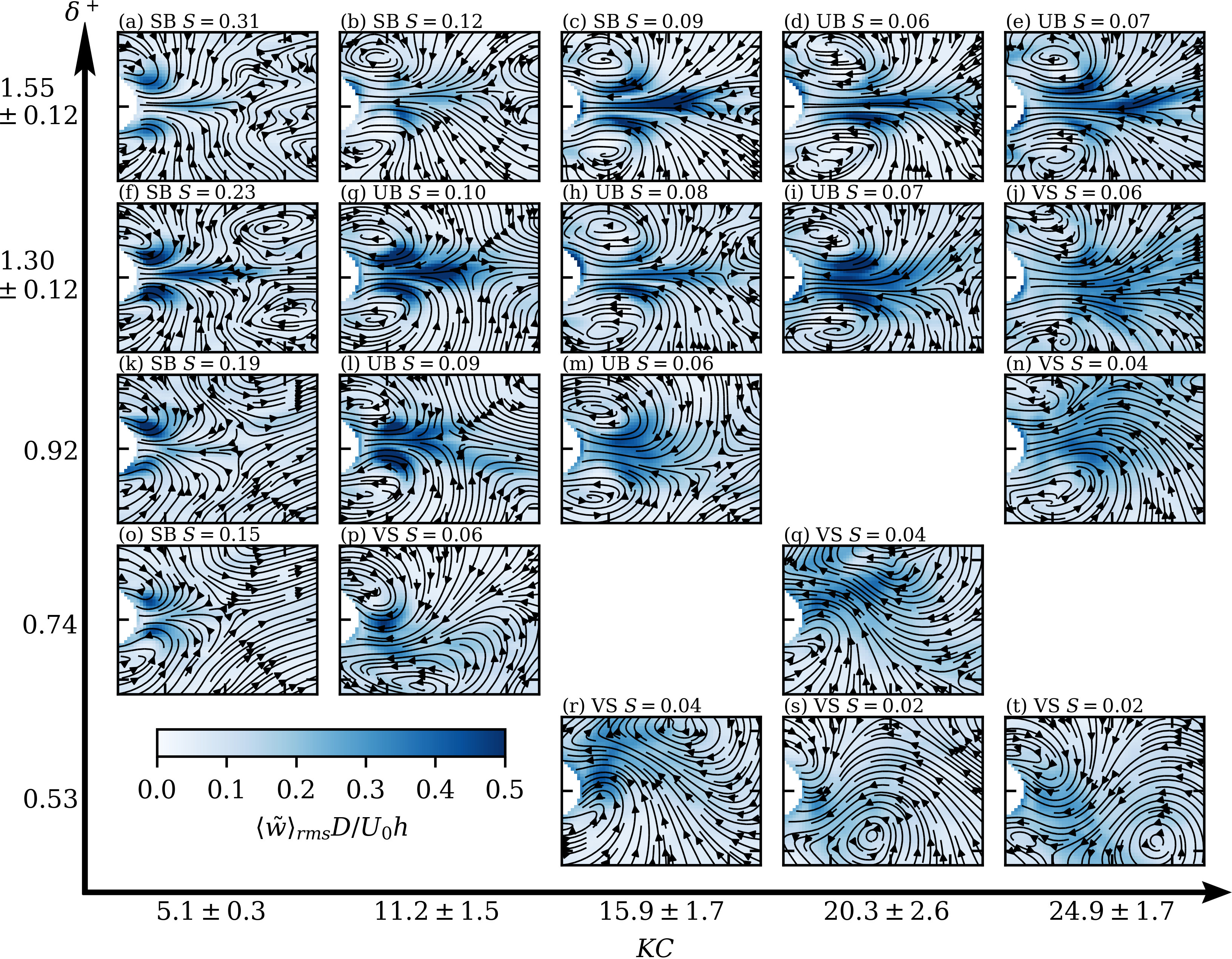}}% Images in 100% size
	\caption{Streamlines of depth- and phase-averaged velocity in the reference frame of the flow. Phase-average calculated over 8
	cycles for a phase of $\phi=0.375$ with shading indicating the depth- and phase-average vertical velocity normalised by $U_0h/D$. Outer axes gives an indication of the approximate variation in $KC$ and $\delta^+$ across the presented runs. Stability increases from bottom right to top left. Inset axes title indicates the wake form (SB, UB and VS) with the field of view and labels as per figure \ref{fig:wakeClassification}.}
	\label{fig:streamLinePlanView}
\end{figure}

According to Equation \ref{eqn:1onS}, the wake stability in shallow oscillatory flow
is governed by the two parameters $KC$ and $\delta^+$. Figure
\ref{fig:streamLinePlanView} provides an overview of the wake forms in this $KC$-$\delta^+$ parameter space. For each experiment the phase and depth averaged
velocity is calculated at a phase of
$\phi=0.375$. Streamlines presented in the frame of reference moving with the flow, at a phase of $\phi=0.375$, provided the clearest visualisation of the wake forms across all wake categories. From the cylinder wake structure of unbounded oscillating flow it is known that
flow separation is initiated at $Re=O(10^2)$ when $KC>3$. For
$Re>O(10^3)$ flow separation occurs at $KC>2$, consistent with the present
results: in figure \ref{fig:streamLinePlanView}o, for $KC=5.0$ and $Re=414$ flow separation is observed.
For $KC<7$, asymmetry commences when $\delta^+<1.0$ (compare figure \ref{fig:streamLinePlanView}f and k). For smaller values of $\delta^+<0.5$, asymmetry is
likely to increase, as is the case in unbounded oscillatory wakes where vortex
asymmetry begins when $KC>4$. In unbounded cylinder wakes vortex shedding commences for $KC>7$ \citep{Sumer2006}. For $KC<7$, the time scale of vortex evolution $D/U_0$ is longer than the period of flow oscillation. This limits the growth of the wake vortices, with equivalent behaviour observed in this study for shallow island wakes.

The spatial distribution of vertical stirring in the wake, relative to the primary vortices visible in the streamline topology, is shown in the normalised, phase- and depth-averaged, vertical velocity (figure
\ref{fig:streamLinePlanView}). 
Upwelling and downwelling is
more localised when the wake bubble remains steady and vortices attached to the island. For increasing $KC$ and $\delta^+\gtrsim 1$ (see figure \ref{fig:streamLinePlanView}f-i) the dominant region of upwelling moves away from the center of the primary wake vortices, towards two elongated secondary vortices (with axes of rotation aligned horizontally). These secondary vortices are associated with a zone flow convergence and divergence along the wake centerline.  
In the vortex
shedding regime, upwelling is increasingly associated with long
secondary vortices that extend laterally between the primary vertical vortices that are shed into the flow. Due to the transient, unsteady character of the secondary vortices, their influence is indicated by the diffuse regions of elevated $\langle \tilde{w} \rangle$ in figure \ref{fig:streamLinePlanView}q,r,s and t.

For given
$\delta^+\lesssim 1.0$ increasing $KC$ increases the number of vortices shed per cycle; however the
number of vortices shed are fewer than unbounded cylinders due to
the influence of bed friction. Further,
increasing $\delta^+$ decreases the shedding frequency (relative to $T$) for a given
$KC$ (the lower limit of which is the complete suppression of vortex shedding)
(figure \ref{fig:streamLinePlanView}d,i,q, and s). 
%In the limit of $\delta^+\ll0.1$ the wake form approaches that predicted
%by $KC$ for an unbounded cylinder.
For a given value of $S$,
for example $S=0.06$ (figure \ref{fig:streamLinePlanView}e, i, o,
and s), wake symmetry increases with $\delta^+$. This highlights the distinct mechanisms that establish wake stability: the relative length of the tide and the relative influence of the boundary layer, even as $S$ remains constant. 

\subsection{Temporal evolution of wake stability}

\begin{figure}
	\centerline{\includegraphics[width=0.9\textwidth]{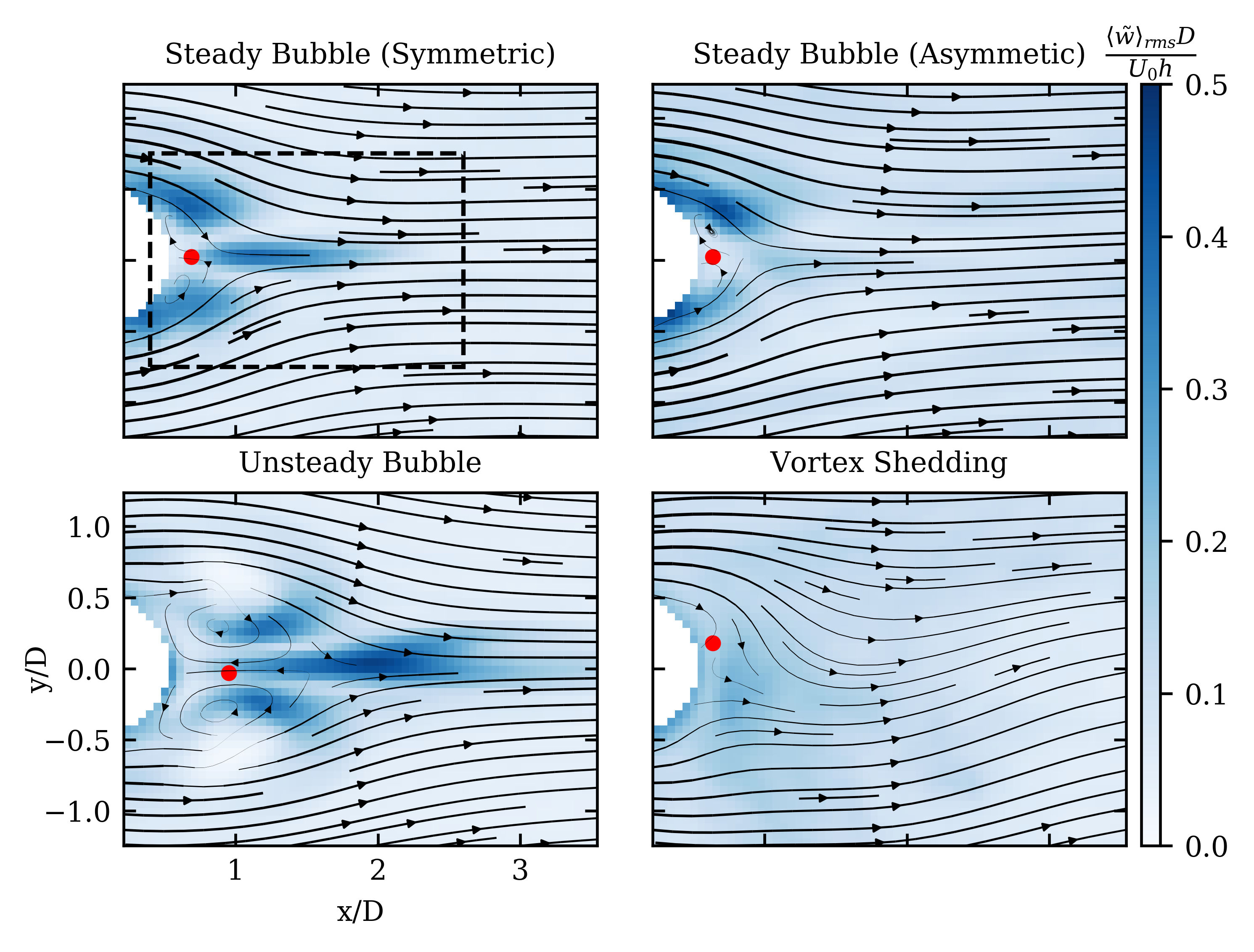}}% Images in 100% size
	\caption{Example calculation of return velocity ($U_m$) for the various classes of wake from phase ($0.0 < \phi < 0.5$) and depth averaged velocity fields. The location of the peak return velocity (the minimum $\langle \tilde{u} \rangle(x,y)$-velocity) is indicated with a red dot. The dashed box indicates the normalised (by $D$) field of view for the larger $D=15$ cm experiments.}
	\label{fig:ReturnFlowExample}
\end{figure}

In steady, shallow wake flow, the magnitude of the return velocity (the flow back towards the cylinder near the wake centreline) is a key indicator of the overall stability of the transverse profile of streamwise velocity \citep{Grubisic1995, Chen1997}. The depth- and phase-averged $u$-velocity is calculated for three different intervals of the phase during the `growth' period. From the resulting  fields of $\langle \tilde{u} \rangle(x,y)$, the return velocity ($U_m$) is identified as the minimum streamwise velocity in the wake. A negative value of $U_m$ indicates a return flow towards the island. Examples of the calculation of the return velocity for the different wake classes are shown in figure \ref{fig:ReturnFlowExample}. 
%The dashed outline indicates the relative size of the measurement field of view for the larger cylinder diameter conditions ($D=15$ cm). 
The red dot indicates the location of $U_m$ (i.e. the reverse flow towards the cylinder) with $x_p$ and $y_p$ defined as the normalised coordinates of the position of $U_m$.  

\begin{figure}
	\centerline{\includegraphics[width=1.0\textwidth]{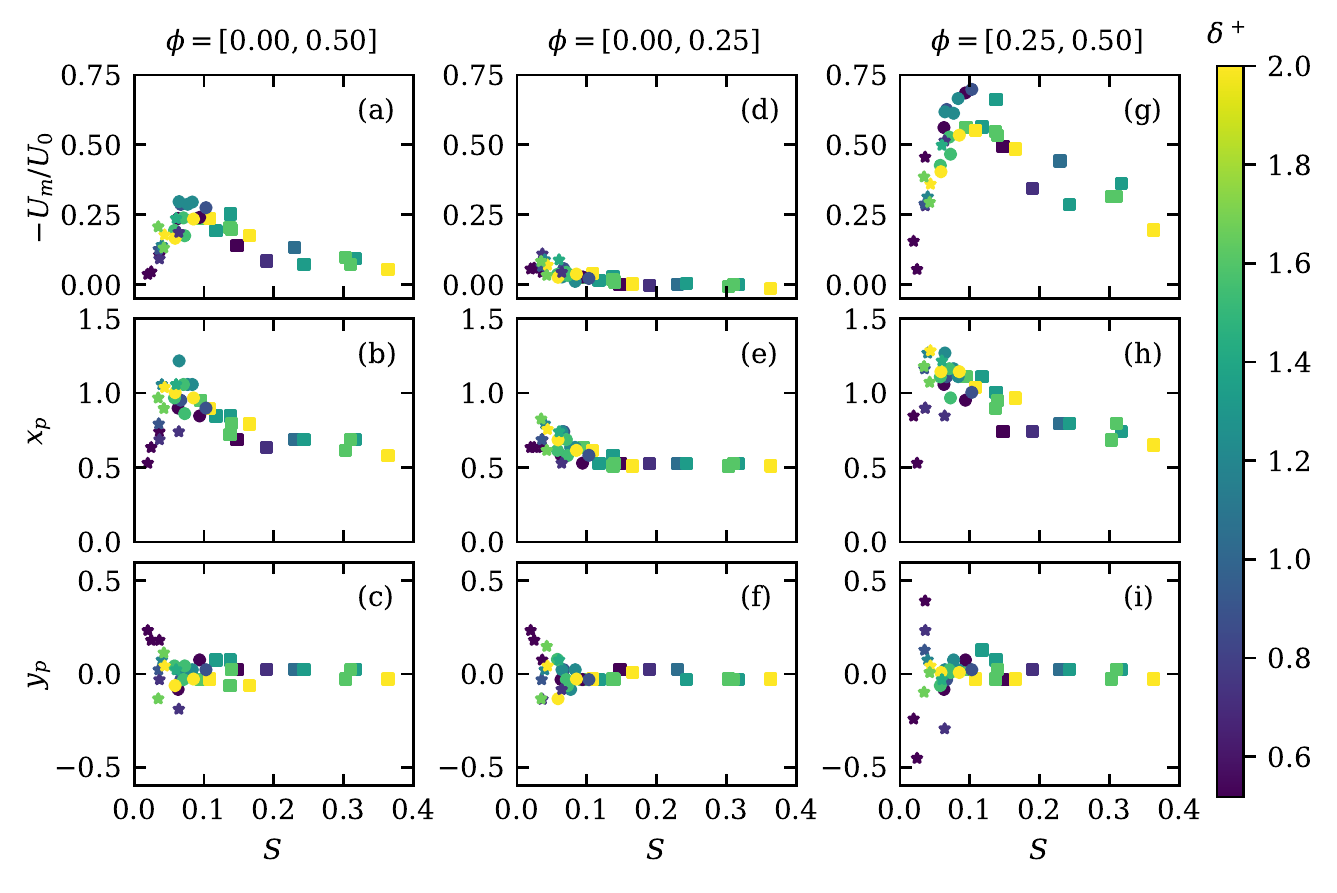}}% Images in 100% size
	\caption{Magnitude ($-U_m/U_0$) and location ($x_p$,$y_p$) of the peak depth and phase-averaged return flow ($U_m$). (a-c) Growth period ($0.0 < \phi < 0.5$), (d-f) accelerating stage of growth period ($0.0 < \phi < 0.25$) and (g-i) decelerating stage of growth period ($0.25 < \phi < 0.5$). Symbols indicate the wake forms as per figure \ref{fig:ParamSpace} and colour indicates the value of $\delta^+$.}
	\label{fig:ReturnFlow}
\end{figure}

The wake stability parameter, $S$, controls the magnitude of the wake return flow (figure \ref{fig:ReturnFlow}). For the growth period, $0.0<\phi<0.5$, the magnitude of the depth-average return flow reaches a maximum of $-0.3U_0$ in the UB wake and decays exponentially with increasing stability (figure \ref{fig:ReturnFlow}a). Similarly, the maximum return flow for the UB wake in the steady-flow experiments of \cite{Chen1995} was $-0.35U_0$, with a rapid drop off for increasing $S$. The peak in the return flow occurs around $x/D\approx 1$ (consistent with the steady-flow wake) and moves closer to the island for increasing wake stability (figure \ref{fig:ReturnFlow}b).  With the onset of vortex shedding for $S\lesssim 0.06$, the magnitude of the phase-averaged return flow reduces sharply. The deflection of the return flow, away from the wake centreline, highlights the asymmetry that results from vortex shedding (figure \ref{fig:ReturnFlow}c).

Vortex shedding wakes have a return flow that is established early in the growth period. The accelerating stage ($0.0 < \phi < 0.25$) of the growth period exhibits a distinct trend of reducing return flow magnitude with increasing stability (figure \ref{fig:ReturnFlow}d), ultimately leading to the complete suppression of the return flow for $S\gtrsim 0.15$. Similarly, the location of the maximum return flow shifts downstream as the strength of the return flow increases (figure \ref{fig:ReturnFlow}e). Vortex shedding commences when the return flow in the wake exceeds approximately $0.05U_0$ during the accelerating stage. The onset of increasing asymmetry, early in the tidal cycle, is demonstrated by the lateral variability of the location of maximum return flow (figure \ref{fig:ReturnFlow}f). 

The dynamics of the wake alter significantly when the pressure gradient imposed by the external flow becomes adverse (for $0.25 < \phi < 0.5$). The return flow induced in the wake is enhanced by the external pressure gradient and a strong return flow jet of up to $0.7U_0$ is present for $S\approx 0.1$ (figure \ref{fig:ReturnFlow}f). The strong return flow leads to earlier flow reversal at the island flanks also with a jet like character. There is a second-order influence on the strength of the return flow due to $\delta^+$, where for a given $S \approx 0.1$, a wake with smaller $\delta^+$ has a stronger return flow (figure \ref{fig:ReturnFlow}g). For $S<0.1$, as the wake stability reduces, the magnitude of the return flow reduces. This is due to the wake centreline region being dominated by lateral flow associated with wake vortices. The lateral location of the return flow can vary widely for the vortex shedding wake at this stage of the tidal cycle (figure \ref{fig:ReturnFlow}i). This lateral variation is a distinguishing feature of the vortex shedding wake, compared to the unsteady bubble wake, which tends to maintain a strong return flow along the wake centreline ($y/D=0$). The location of the maximum return flow is furthest downstream at the transition to vortex shedding, $S\approx 0.05-0.07$ (figure \ref{fig:ReturnFlow}h), even though the maximum return flow occurs at $S\approx 0.1$.  

In conjunction with observations of the wake form, the magnitude of the return flow at different stages of the tide gives an indication of the global stability of the wake circulation. The maximum growth rate of the absolute instability in the wake is controlled by the strength of the return flow with the downstream extent of the region of return flow indicative of the size of the region of absolute instability \citep{Grubisic1995}. The fact that the location of the return flow is furthest from the cylinder at the transition to vortex shedding supports the concept that a region of absolute instability must grow sufficiently large within the tidal cycle for the onset of vortex shedding. Furthermore, when the external flow is unsteady, a return flow of sufficient strength must be establish during the accelerating stage of the tide for the onset of vortex shedding. These results demonstate that consideration of the tidal phase is crucial in interpreting observations of island wakes in tidal flow.

\subsection{Scaling of lateral and vertical velocity}

A novel and significant contribution of this study is the full 3D velocity field
measured over a wide range of parameters. This provides a solid basis for the
scaling of lateral and vertical (upwelling) velocities in the tidally-forced island wake. For shallow flow, where bottom friction is important, upwelling in the island wake is expected to be driven by the curl of stress at the bed associated with the wake eddies. A simple Ekman pumping model is a reasonable approach to examine the scaling of vertical velocity. The Ekman velocity is \citep{CushmanRoisin2011}:
\begin{equation}
	w_E = \frac{1}{\rho}\overrightarrow{k}\cdot\nabla\times(\tau_b/\omega_e)
	\label{eqn:We}
\end{equation}
where $\overrightarrow{k}$ is the vertical unit vector, $\tau_b$ 
is the bed shear stress and $\omega_e$ is a characteristic eddy rotational 
frequency. The bed stress will scale as $\tau_b\sim\mu U_e/\delta_E$, where $\delta_E$ is the thickness of the Ekman boundary layer and $U_e$ is a characteristic azimuthal velocity for the wake eddies. The eddy rotational frequency will scale as $\omega_e\sim U_e/L_e$, where $L_e$ is a characteristic length scale of the wake eddies. The Ekman boundary layer thickness will scale as $\delta_E\sim\sqrt{\nu/\omega_e}$. Letting the gradient scale as $\nabla \sim 1/L_e$ and substituting into equation (\ref{eqn:We}) gives:
\begin{equation}
	w_E \sim \sqrt{\frac{\nu U_e}{L_e}}
	\label{eqn:We2}
\end{equation}
where $U_e$ and $L_e$ have an unknown functional dependence on the parameters $h/D$, $KC$ and $\delta^+$, such that:
\begin{equation}
	U_e/U_0 \sim f_U(h/D,KC,\delta^+) = \beta _{0,f_U}(h/D)^{\beta _{h/D,f_U}}KC^{\beta _{KC,f_U}}(\delta^+)^{\beta _{\delta^+,f_U}}
	\label{eqn:fu}
\end{equation}
and 
\begin{equation}
	L_e/D \sim f_L(h/D,KC,\delta^+) = \beta _{0,f_L}(h/D)^{\beta _{h/D,f_L}}KC^{\beta _{KC,f_L}}(\delta^+)^{\beta _{\delta^+,f_L}}
	\label{eqn:fl}
\end{equation}
where $\beta _{p,r}$ are the scaling coefficients for parameter $p$ and scaling relationship $r$.

Substituting (\ref{eqn:fu}) and (\ref{eqn:fl}) in (\ref{eqn:We2}) gives
\begin{equation}
	w_E/U_0 \sim \frac{1}{\sqrt{Re}}\sqrt{\frac{f_U}{f_L}} \equiv \frac{h}{D}\frac{\delta^+}{\sqrt{KC}}\sqrt{\frac{f_U}{f_L}}
	\label{eqn:wU01}
\end{equation}
where the scaling coefficients of $f_U$ and $f_L$ can be estimated from three-dimensional velocity measurements of the shallow island wake assuming that $U_e \sim \langle \overline{v} \rangle$ and $w_E \sim \langle \overline{w} \rangle$.

\begin{figure}
	\centerline{
		\includegraphics[width=0.95\textwidth]{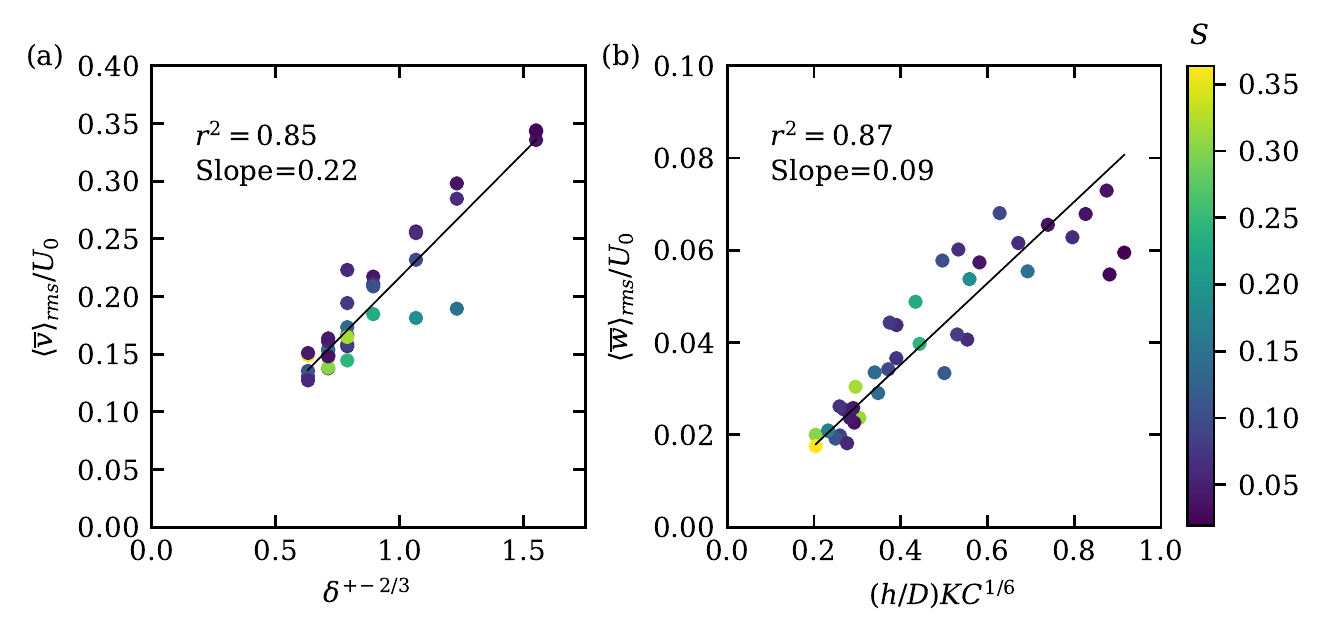}
	}% Images in 100% size
	\caption{Scaling relationships for (a) $\langle \overline{v} \rangle_{rms}/U_0$ and (b) $\langle \overline{w} \rangle_{rms}/U_0$ based on a simple parametrisation of $U_e/U_0\sim f_U(\delta^+)$ and $L_e/D\sim f_L(KC)$. The black line indicates a 1:1 slope.}
	\label{fig:ScalingSim}
\end{figure}

The Bayesian statistical analysis (see Supplementary Material) indicates
that there is a range of credible
scaling coefficient values for $f_U$ and $f_L$ that can adequately describe the relationship
between the eddy velocity and length scales that drive a vertical flux from the boundary
layer via Ekman pumping.
The posterior distribution of the scaling coefficients indicates that the effect size of the flow aspect ratio ($h/D$) in $f_U$ and $f_L$ is likely zero. Examination of the remaining scaling coefficients (for $KC$ and $\delta^+$) indicates that $\beta _{KC,f_L} \approx -\beta _{\delta^+,f_L}$ which implies that $L_e \sim D.S^{\beta _{\delta^+,f_L}}$. Taking $\beta _{KC,f_U}=0$, $\beta _{\delta^+,f_U}=-2/3$, $\beta _{KC,f_L}=-4/3$ and $\beta _{\delta^+,f_L}=4/3$ as a particular model within the 95\% confidence intervals of the scaling coefficients gives:
\begin{equation}
	\langle \overline{v} \rangle_{rms}/U_0=0.22(\delta^+)^{-2/3}
	\label{eqn:vSim}
\end{equation}
\begin{equation}
	\langle \overline{w} \rangle_{rms}/U_0=0.09(h/D)KC^{1/6}
	\label{eqn:wSim}
\end{equation}
which predicts well the first order scaling of $\langle \overline{v} \rangle_{rms}$ ($r^2=0.85$) and $\langle \overline{w} \rangle_{rms}$ ($r^2=0.87$) over the range of these experiments (figure \ref{fig:ScalingSim}). This demonstrates that in the island wake, the lateral velocity scales with the relative tidal boundary layer thickness, due to entrainment of the vertical structure of the external forcing flow into the wake. The upwelling velocity scales predominantly with the flow aspect ratio (as would be expected by conservation of mass) with a weak increase for increasing relative tidal excursion. 

A number of observations can be drawn from the implied scaling relationships: 
\begin{equation}
	L_e \sim DS^{4/3}
\end{equation}
\begin{equation}
	\omega_e \sim \frac{U_0}{D} \frac{KC^{4/3}}{(\delta^+)^2} \equiv \frac{U_0}{D} \frac{1}{S^{4/3}(\delta^+)^{2/3}}
\end{equation}
\begin{equation}
	\delta_E \sim h\frac{(\delta^+)^2}{KC^{7/6}} \equiv h S^{7/6} (\delta^+)^{5/6}.
\end{equation}
Firstly, the characteristic length scale of the wake eddies scales with the island diameter and increases with increasing wake stability, consistent with observations of wake eddies presented \citep{Branson2018a}. Secondly, the characteristic time scale of the wake eddies ($1/\omega_e$) scales with $D/U_0$ (consistent with the unbounded cylinder wake) and increases with increasing stability and relative boundary layer thickness whilst decreasing with increasing relative tidal length. Finally, the wake Ekman boundary layer scales with the flow depth and increases with increasing stability, and relative tidal boundary layer thickness while decreasing with increasing relative tidal length (as expected with the decrease in characteristic eddy time scale).

\subsection{Distribution of vertical velocity}

\begin{figure}
	\includegraphics[width=1.0\textwidth]{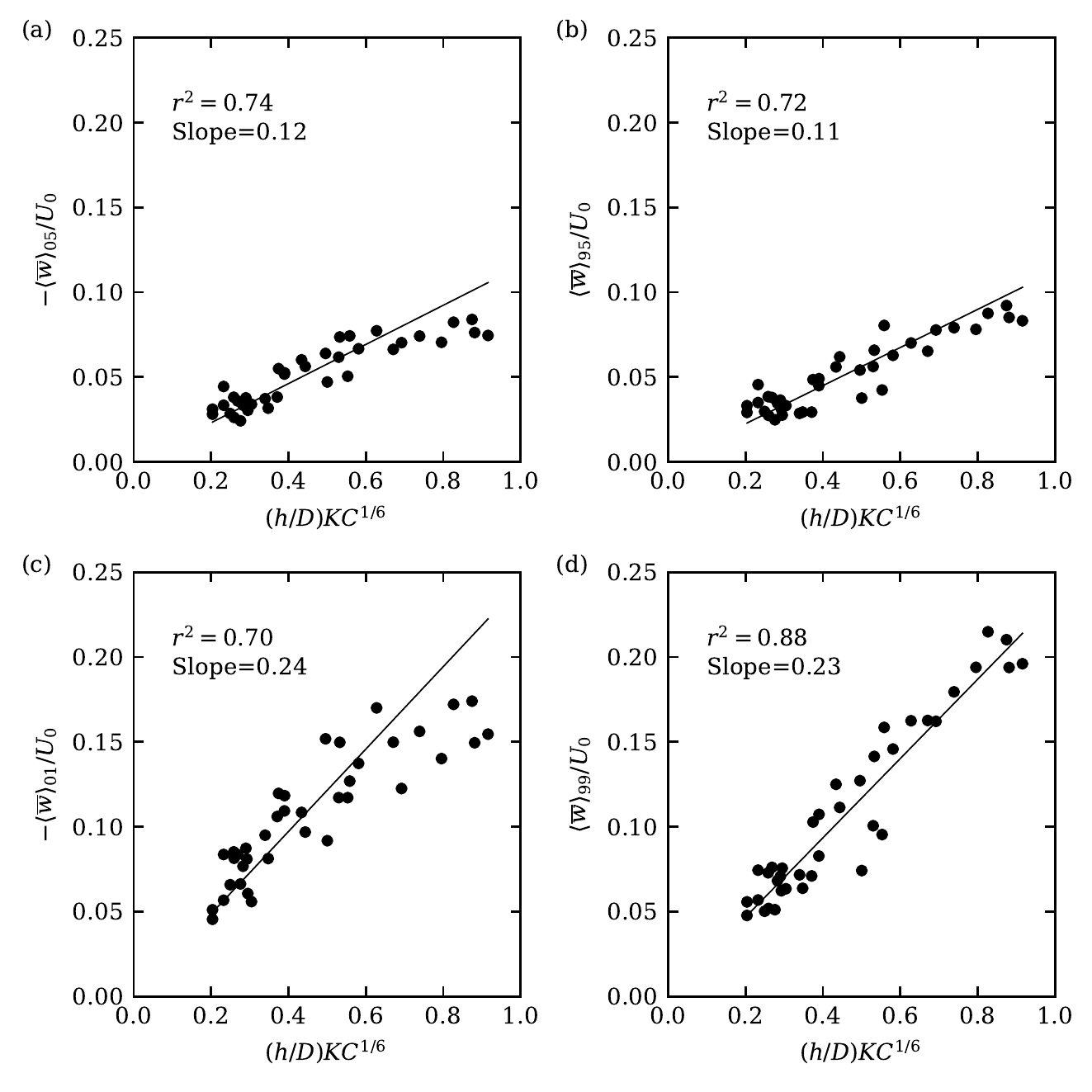}
	\caption{Scaling relationships for the (a) fifth (downwelling) and (b) ninety-fifth percentile (upwelling) velocities and (c) first (downwelling) and (d) ninety-ninth percentile (upwelling) velocities. The black line represents the model estimate with the indicated slope.}
	\label{fig:ScalingPeaks}
\end{figure}

For $(h/D)KC^{1/6}\gtrsim 0.6$, a second-order influence on the scaling of the vertical velocity is suggested by a trend in the measured data away from the modelled line (figure \ref{fig:ScalingSim}). It is possible to examine skewness in the distribution of vertical velocity by examining the scaling of the peaks in the up- and downwelling velocity (which will have a strong influence on the rms). The vertical velocity scaling relationship given in Equation \ref{eqn:wSim} is robust for the peaks in the up- and downwelling velocity (figure \ref{fig:ScalingPeaks}). The distribution between up- and downwelling is reasonably symmetric, as indicated by the similar slope for the first, and ninety-ninth percentiles ($\approx0.24$) and, the fifth, and ninety-fifth percentiles ($\approx0.12$). The deviation with increasing $(h/D)KC^{1/6}\gtrsim 0.6$ is greatest for the peaks in downwelling and not present for the peaks in upwelling. The dominant region of upwelling is in the centre of the primary wake eddies, where rotation is greatest. Downwelling tends to be associated with secondary vortices, in zones of flow convergence, strain, and vortex stretching. The dynamics of the downwelling regions are most significantly influenced by the reduced flow stability associated with increasing relative tidal excursion. If the magnitude of the peaks in downward flux are reduced compared to upward flux, the integrated area of downward flux must also be larger (by conservation of mass). 

\section{Discussion}
\begin{figure}
	\centerline{
		\includegraphics[width=1.0\textwidth]{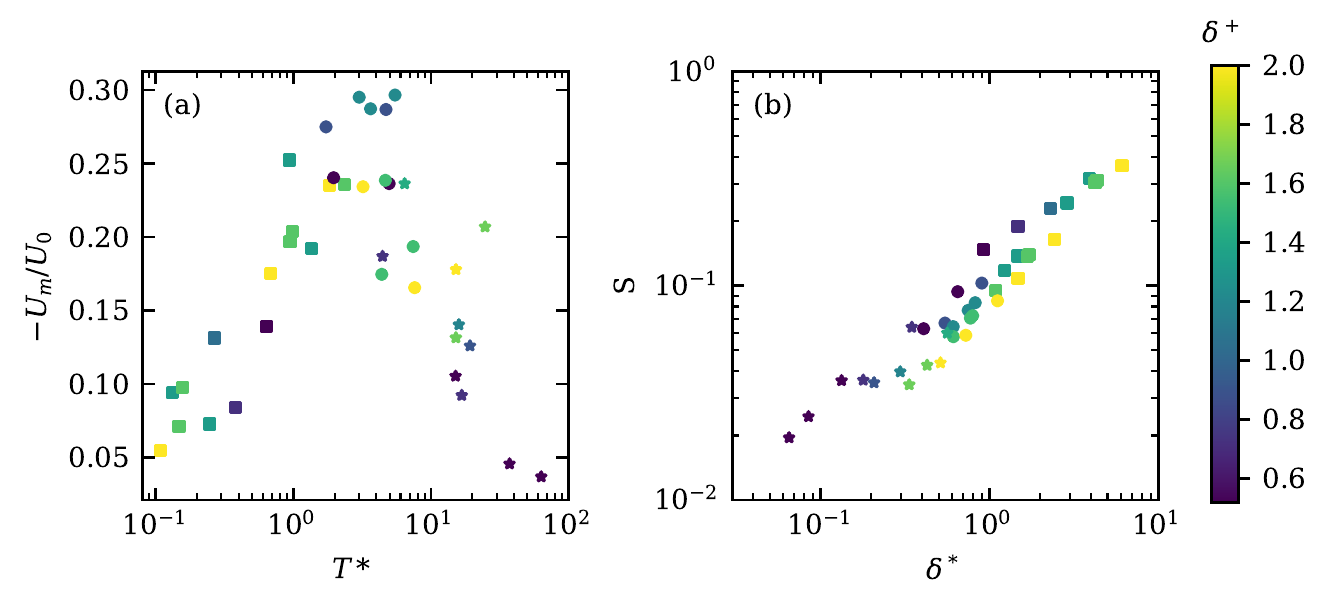}
	}% Images in 100% size
	\caption{(a) Variation of maximum wake return flow $-U_m/U_0$ with the relative eddy revolution time scale $T^*$ and (b) the wake stability parameter $S$ with the relative Ekman boundary layer thickness $\delta^*$. Symbols indicate the wake forms as per figure \ref{fig:wakeClassification} and colour indicates the value of $\delta^+$. Transitions in the wake dynamics (indicated by the return flow) and wake form are predicted well by the implied eddy scales calculated from a model of wake upwelling driven by Ekman pumping.}
	\label{fig:JFM_EddyScales}
\end{figure}

Vortex shedding is a mechanism by which wake eddy circulation is released from the low-pressure, wake bubble region. Prior to the onset of vortex shedding we would expect the magnitude of the return flow $U_m$ to be a maximum when the tidal period $T$ is close to the time scale for a wake eddy to complete a revolution (i.e. through a resonance-like mechanism). Taking $U_e=0.22U_0\delta^{+\!-2/3}$, the upwelling velocity $w_E=\sqrt{2\nu\omega_e}=0.09U_0(h/D)KC^{1/6}$, and the eddy azimuthal frequency $\omega_e=U_e/L_e$ with $L_e=c_LDS^{4/3}$ where $c_L$ is a constant, gives $c_L\approx1.76$. The implied time for a wake eddy to complete a revolution is $T_e=2\pi L_e/U_e$ from which we can calculate the relative length of the tidal period to the eddy revolution time as $T^*=T/T_e$. The magnitude of the return flow increases as $T^*$ increases to 1 (figure \ref{fig:JFM_EddyScales}a). A transition in wake form, to the unsteady bubble wake, occurs for $T^*\approx2$. This stands to reason given the wake eddies are generated for each half-cycle of the tide. Finally, vortex shedding commences for $T^*\gtrsim7$, demonstrating that $T*$ in shallow flow is analogous to $KC$ in unbounded oscillatory flow.

It is also possible to examine the implied relative length scale of the Ekman boundary layer compared to the flow depth. Taking $\delta_E=\sqrt{2\nu/\omega_e}$, with the relative Ekman boundary layer thickness given as $\delta^*=2\pi\delta_E/h$, we see that $S$ in fact predicts the first-order change in relative Ekman boundary layer thickness (figure \ref{fig:JFM_EddyScales}b). At $S=0.1$, $\delta^*=1$ which once again marks the transition to the UB wake. For $\delta^*\lesssim 0.5$ vortex shedding commences. The second order effect of $\delta^+$ on the wake stability is also nicely demonstrated. For fixed $S$, increasing $\delta^+$ increases the Ekman boundary layer thickness (with the converse true for $KC$ by definition - see equation \ref{eqn:1onS}). This explains the variation in wake character shown in figure \ref{fig:streamLinePlanView} for fixed values of $S$. When the relative Ekman boundary layer is smaller (for given $S$), greater asymmetry is observed in the wake. 

Importantly, these results support a model of island wake upwelling driven by Ekman pumping. It highlights that the global wake stability is intimately linked to the evolution of the wake eddies. The wake eddies grow, entraining the vertical structure of the external tidal flow across the separation shear layers at the island flanks. Disturbances introduced into the wake (sourced likely from the separated shear layers) are insufficiently damped when the eddy revolution time scale is shorter than the tidal period and the Ekman boundary layer thickness is smaller than the flow depth. This leads to growing asymmetry in the wake, with an increasingly strong return flow jet that is strongest during the decelerating stage of the tide. The return flow jet leads to earlier flow separation and a jet at the island flanks in the subsequent cycle. These jets, present at the island flanks during flow reversal, were observed to eject circulation away from the island. This appeared to be part of the mechanism by which the establishment of vortex shedding was suppressed for the unsteady bubble wakes that were close to vortex shedding. The wake (and shear layer separation) is convectively unstable, however excitation of a global mode of vortex shedding has not occurred. These intricate details of the half-cycle interactions for incipient vortex shedding is an interesting area of further research. 

Vortex shedding commences if the wake eddies are able to attain sufficient circulation during the accelerating phase of the tide. This is indicated by the strength of the return flow established during the accelerating stage of the tide. At the transition to vortex shedding, one of the wake eddies attains sufficient circulation (and associated low pressure) to steer the return flow preferentially around a particular flank of the island. The return flow, in turn, causes the larger eddy to convect away from the island during flow reversal, releasing the circulation from the island wake generation zone. Close to the transition to vortex shedding this process may not occur every cycle. A lower-frequency, beating-like modulation was observed for cases close to vortex shedding, where an eddy may be shed every few cycles. 

The experiments of this study were at lower $Re$ (and thus smaller $\beta=Re/KC$) than those of \cite{Lloyd2001}. For a given combination of $KC$ and $\delta^+$ (that establish $S$) it is expected that $\beta$ will have an influence by altering the entrainment across the lateral shear layers that extend from the island flanks. We investigated the sensitivity of the wake return flow to $\beta$ for $\delta^+\gtrsim1.4$ 
and found minimal influence with an increase of $\beta$ from 85 to 190. It is possible that the conflicting wake classifications (between this study and \cite{Lloyd2001}) at the transitions between wake forms are due to some sensitivity to $\beta$ (see figure \ref{fig:wakeClassification}). The independent influence of $\beta$ is an area for future work.

\section{Application of laboratory results to island wakes}

The knowledge gained from the laboratory study gives insight into expected behaviour in the field and, in particular, can be used to design field measurement and observation programs. The laboratory results identified the parameters that govern island wakes in tidal flow: $h/D$, $KC$ and $\delta^+$ (with the wake stability parameter $S$ dependent on the ratio of $\delta^+$ and $KC$). The aspect ratio $h/D$ and the Keulegan-Carpenter number $KC$ can be estimated from bathymetric survey, field measurements and knowledge of the local tidal characteristics. 

Whilst in the laboratory the forcing flow was laminar (with molecular viscosity), in the field knowledge of the equivalent vertical eddy diffusivity of momentum $\nu_T$ is needed. 
The vertical transfer of momentum is given by $\tau=\rho\nu_T\partial u/\partial z$, where $\tau$ is the total stress, and $\rho$ the fluid density. When the flow is steady and shallow, with a logarithmic boundary layer that extends over the fluid depth, $\nu_T\sim u_*h$ where $u_*=\sqrt{\tau_b/\rho}$ is the shear velocity \citep{Fischer1973}. When the flow is unsteady and deep, the transfer of momentum to the bed occurs across a turbulent bottom boundary layer with a characteristic length scale given by $\delta\sim u_*/\omega$ and hence $\nu_T\sim u_{*}\delta$ (see, e.g., \cite{Jensen1989} and \cite{CushmanRoisin2011}). It is the transition from $\nu_T\sim u_{*}\delta$ to $\nu_T\sim u_*h$ with decreasing flow depth which must be resolved for tidal flows on continental shelves and warrants further investigation, particularly when $\delta\approx h$. When $\delta\approx h$ the length scale of the largest eddy that transports momentum vertically will be influenced by both the flow depth and tidal period. To allow estimation of $\delta^+$, it is clear that future field data collection exercises must seek to measure both the vertical structure of the mean horizontal velocity and turbulent stress of the incoming tidal flow. In addition, measurements of the vertical structure of the horizontal velocity, and turbulent stress in the wake return flow, will assist in evaluating the structure of the wake eddies and influence of tidal flow unsteadiness. By way of example, we consider two field sites where island wakes have been observed.

\subsection{Observations of Rattray Island, Australian Great Barrier Reef}

It is informative to evaluate the results of this study against field observations of island wakes in tidal flow. In this context, Rattray Island in the Australian Great Barrier Reef is one of the best studied, with detailed field observations presented in \cite{Wolanski1984} and additional analyses in \cite{Wolanski1996,Wolanski2003}. 

\cite{Wolanski1996} estimated an upwelling velocity of $0.002-0.004$ m~s$^{-1}$ based on the time it takes suspended mud to reach the surface in an eddy. The semi-diurnal tidal velocity during the study of Rattray Island was reported as 0.7 m~s$^{-1}$, with an island size of 1500 m. This gives $KC\approx20$. The scaling laws presented in figure \ref{fig:ScalingSim} and \ref{fig:ScalingPeaks} give a predicted rms and peak (95$^{th}$ percentile) upwelling velocity of 0.002 and 0.003 m~s$^{-1}$, respectively, in good agreement with field estimates.

In terms of the return flow, \cite{Wolanski2003} reported that the maximum velocity in the wake eddy at Rattray island was $0.5-0.75U_0$ occuring approximately 1 hour after peak tidal flow \citep{Wolanski1996}.  Previous laboratory investigations of shallow island wakes in a steady external flow have reported an upper limit on the maximum return flow velocity of approximately $0.35U_0$ for unsteady bubble wakes \citep{Chen1995,Wolanski1996}. In this study, consistent with the field observations, a maximum return flow of $0.5-0.75U_0$ occurred after peak tidal flow for unsteady bubble wakes with $S\approx 0.1$ (figure \ref{fig:ReturnFlow}). \cite{Wolanski1984} assumed a quadratic bottom friction coefficient of $c_f = 2.5 \times 10^3$, which from $u_*=\sqrt{c_f/2}U_0$ gives a friction velocity of $u_*=2.5 \times 10^{-2}$ ms$^-1$. Assuming that $\delta=0.4u_*/\omega$ \citep{CushmanRoisin2011} gives $\delta/h\approx2.7$ and $S\approx0.14$, which is reasonably consistent with the observed magnitude of the return flow. This magnitude of return flow occurs due to the adverse external pressure gradient present after peak tidal flow and highlights the importance of incorporating flow unsteadiness in the study of shallow island wakes subjected to tidal forcing. 

\cite{Wolanski1984} estimated the turbulent horizontal eddy viscosity of the flow around Rattray Island as $O(1$ m$^2$s$^{-1}$) by releasing groups of surface floating drogues. This suggests a Reynolds number (using the eddy viscosity) of $O(1000)$ and $\beta=Re/KC$ of $O(100)$. This indicates that the values of $\beta$ achieved in this laboratory study $(50-200)$ are comparable to the field range when an appropriate turbulent horizontal eddy viscosity is utilised (noting that the turbulent eddy viscosity is a function of the flow and will vary in space and time).

\subsection{Observations of the wakes of islands in the Kimberley region of north western Australia}

\begin{figure}
	\centering
    \includegraphics[width=0.75\textwidth]{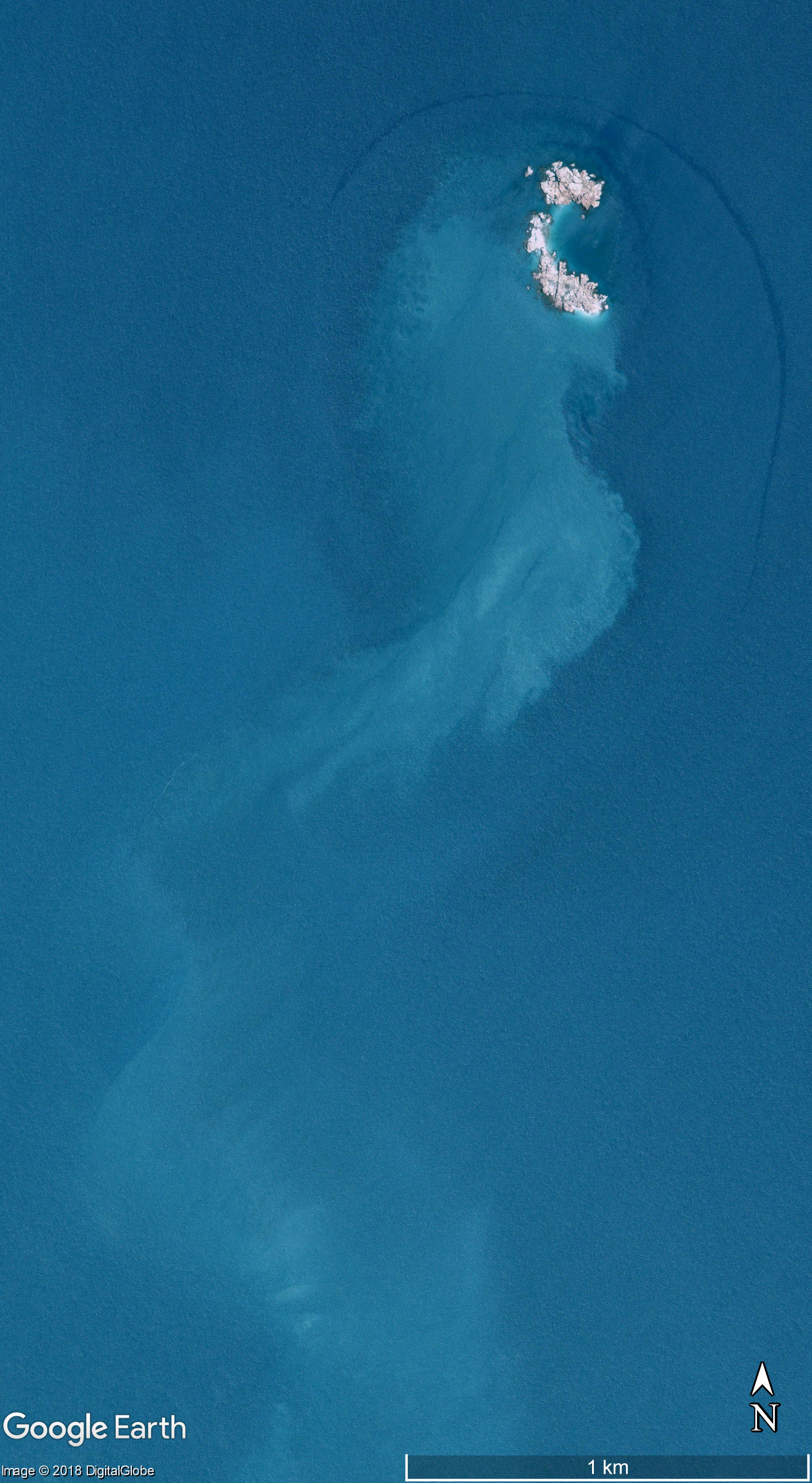}
	\caption{Satellite photo of an island in the Kimberley Region (14.46$^{\circ}$S, 125.32$^{\circ}$E) on 19 March 2003 showing a vortex shedding wake. The darker arc indicates a region where the sub-surface flow is influencing the free surface roughness. From Google Earth with image \copyright~2018 DigitalGlobe.}
	\label{fig:KimberlayWakeZoom}
\end{figure}

The laboratory experiments in this thesis were undertaken with cylindrical islands that have vertical side walls. 
In the Kimberley region of north western Australia, the islands can have very steep sides due to the geomorphology of the region. In addition, the tidal range is very large, $O(10)$ m, with tidal currents that may exceed 2 m~s$^{-1}$ \citep{Creswell2000}. During field data collection in the Kimberley region, large free surface perturbations associated with tidal eddies were observed (by the author), suggesting that non-hydrostatic processes may also be relevant at field scales. Remote sensing images also support this. Figure \ref{fig:KimberlayWakeZoom} shows a vortex shedding wake in the Kimberley region of Western Australia. The arc that outlines the northern side of the island indicates a sub-surface flow structure that is influencing the free surface (suggestive of the presence of a horseshoe vortex). This island is approximately 300 m in diameter with tidal currents assumed to be $O(2$ ms$^{-1})$ which gives $KC=O(200+)$. A recently developed bathymetric grid for northern Australia indicates that the island is present in water approximately 20 m depth. Assuming $u_*=O(0.1$ ms$^{-1})$ gives $\delta=O(200$ m) and $S=O(0.03)$,  suggesting an extensive vortex shedding wake which is consistent with figure \ref{fig:KimberlayWakeZoom}. Remote sensing offers significant opportunities for the further study of shallow island wakes, with a recent study demonstrating the use of along track stereo images from the WorldView-3 satellite to obtain surface velocity measurements \citep{Delandmeter2017}.

\section{Conclusion}

Three classes of wake form have been identified in the tidally-forced, shallow island wake: \emph{(1) steady bubble, (2) unsteady bubble} and \emph{(3) vortex shedding}. The two parameters $KC$ and $\delta^+$ combine to establish the island wake stability $S$ for shallow tidal flow. The \emph{unsteady bubble} wake is observed for $S\lesssim 0.1$ with the onset of the \emph{vortex shedding} wake for $S\lesssim 0.06$. The unsteady tidal flow establishes two distinct phases in the evolution of the shallow island wake, the accelerating stage, prior to peak tidal flow, and the decelerating stage where the external pressure gradient is adverse. Vortex shedding is only observed if a return flow region of sufficient size and strength is established during the accelerating stage.  

Through the novel experimental technique applied in this study, the three-dimensional flow field of the unsteady, shallow island wake has been resolved. A model of wake upwelling based on Ekman pumping from the bottom boundary layer is established. Lateral (cross-stream) velocity in the wake scales primarily with $U_0(\delta^{+})^{-2/3}$ and is demonstrated to be an effective proxy for the wake eddy azimuthal velocity. Vertical velocity scales with $U_0(h/D)KC^{1/6}$ across a range of statistical measures of the mean and peak up- and downwelling velocity. Deviation from the scaling relationships is observed for $(h/D)KC^{1/6} \gtrsim 0.6$, which is primarily associated with a reduction in the peak downwelling velocities. To lowest order, the upwelling magnitude is independent of the wake form, with the wake form primarily influencing the spatial distribution of the up- and downwelling regions relative to the island. The simple Ekman pumping model predicts the vertical transport for wake forms that have significantly different topologies of primary and secondary vortices.

Finally, an intrinsic link is demonstrated between the implied dynamical scales of the wake eddies (predicted by the Ekman pumping model) and the wake stability. The wake return flow is strongest when the time scale of wake eddy revolution is the same order as the tidal period. The stability parameter $S$ predicts the relative thickness of the Ekman boundary layer to flow depth, with a secondary dependence on the relative value of $\delta^+$ and $KC$.   

\begin{itemize}
    \item Funding was provided by Australian Research Council Discovery
    Project (Grant No. DP1095294).
    \item This work was supported by resources provided by The Pawsey Supercomputing
    Centre with funding from the Australian Government and the Government of Western
    Australia. 
    \item E. J. Hopfinger acknowledges a Gledden Visiting Fellowship during his
    stay at The University of Western Australia. 
    \item The authors wish to thank M. E. Negretti for helpful comments and suggestions on a first draft of the paper.  
\end{itemize}

\bibliographystyle{jfm}
\bibliography{paper}

\end{document}

% \maketitle
\newpage

{\centering \sloppy
 {\normalfont\LARGE\fontswitch\bfseries{Supplementary Material}}}

\section{Statistical analysis of power scaling laws}

In log-space the scaling relationships for the eddy azimuthal velocity and eddy length scale establish a general linear system for which the scaling parameters can be estimated using ordinary linear regression or Bayesian statistical methods. The general linear system for the eddy velocity is
\begin{equation}
	\log(\mu_{U}) = \beta _{0,f_U} + \beta _{h/D,f_U} \log(h/D) + \beta  _{KC,f_U} \log(KC) + \beta _{\delta^+,f_U} \log(\delta^+) + \epsilon_U
\end{equation}
and for the eddy length scale is
\begin{equation}
	\log(\mu_{L}) = \beta _{0,L} + \beta _{h/D,f_L} \log(h/D) + \beta  _{KC,f_L} \log(KC) + \beta _{\delta^+,f_L} \log(\delta^+) + \epsilon_L
\end{equation}
where $\mu_U$ and $\mu_L$ are the expected value of the eddy velocity and eddy length scale and $\beta _p,r$ are the scaling coefficients for $p$ variables and $r$ indicating the eddy length ($f_L$) or eddy velocity ($f_U$) scaling relationship. Assuming that the eddy velocity scales $U_e \sim \langle \overline{v} \rangle _{rms}$, we can use $\langle \overline{v} \rangle _{rms}$ and $\langle \overline{w} \rangle _{rms}$ as two readily measurable quantities from which to estimate the scaling coefficients of $f_U$ and $f_L$. 
%We examine the posterior probability distributions of the scaling parameters to establish the minimum set of parameters required to estimate $f_U$ and $f_L$ and their likely ranges.   

\begin{table}
	\begin{center}
		\def~{\hphantom{0}} 
		\makebox[1.0\textwidth][c]{
		\begin{tabular}{lcccccc}
			\toprule
			&        \multicolumn{3}{c}{OLS} & \multicolumn{3}{c}{HMC} \\
			\cmidrule(lr){2-4}\cmidrule(lr){5-7}
			& 95\% LCL & MLE & 95\% UCL & 95\% LCL & Mean & 95\% UCL \\
			\midrule
			$\beta_{0,f_U}$ & -0.83 & -0.75 & -0.66 & -0.85 & -0.75 & -0.65  \\
			$\beta_{h/D,f_U}$ & -0.16 & 0.01 & 0.18 & -0.18 & 0.01 & 0.20  \\
			$\beta_{KC,f_U}$ & 0.02 & 0.07 & 0.12 & 0.02 & 0.07 & 0.13  \\
			$\beta_{\delta^+,f_U}$ & -0.87 & -0.67 & -0.47 & -0.90 & -0.67 & -0.45  \\
			$\beta_{0,f_L}$ & 0.93 & 1.22 & 1.50 & 0.90 & 1.21 & 1.54  \\
			$\beta_{h/D,f_L}$ & -0.42 & 0.16 & 0.74 & -0.50 & 0.16 & 0.79  \\
			$\beta_{KC,f_L}$ & -1.30 & -1.14 & -0.97 & -1.33 & -1.14 & -0.96  \\
			$\beta_{\delta^+,f_L}$ & 0.47 & 1.14 & 1.81 & 0.37 & 1.14 & 1.88  \\
			\bottomrule
		\end{tabular}
		}
\caption{Estimates of scaling coefficients of $f_U$ and $f_L$. Maximum likelihood estimate (MLE) (from ordinary least squares - OLS) and the zeroth moment of the posterior probability distribution (from Hamiltonian Monte Carlo - HMC). The 95\% lower confidence level (LCL) and 95\% upper confidence level of the coefficient estimates. This model has a coefficient of determination of $r^2=0.88$ for both $\langle \overline{v} \rangle _{rms}/U_0$ and $\langle \overline{w} \rangle _{rms}/U_0$.
}
\label{tab:MLE}
\end{center}
\end{table}

\begin{figure}
	\centerline{
		\includegraphics[width=0.5\textwidth]{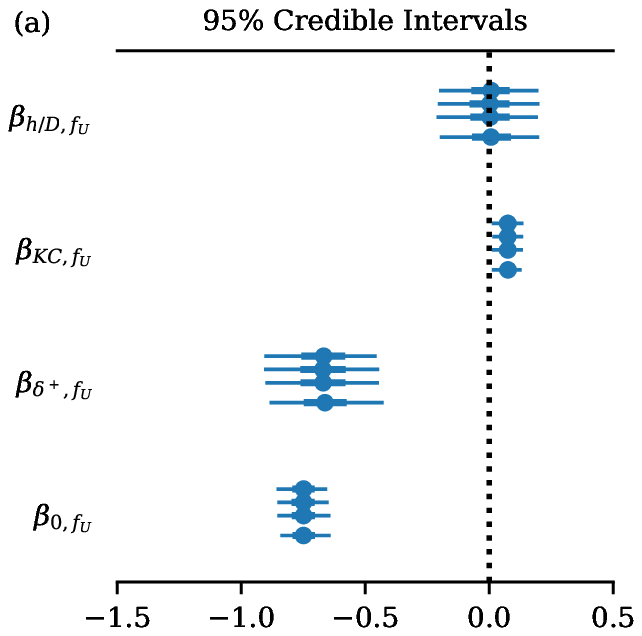}
		\includegraphics[width=0.5\textwidth]{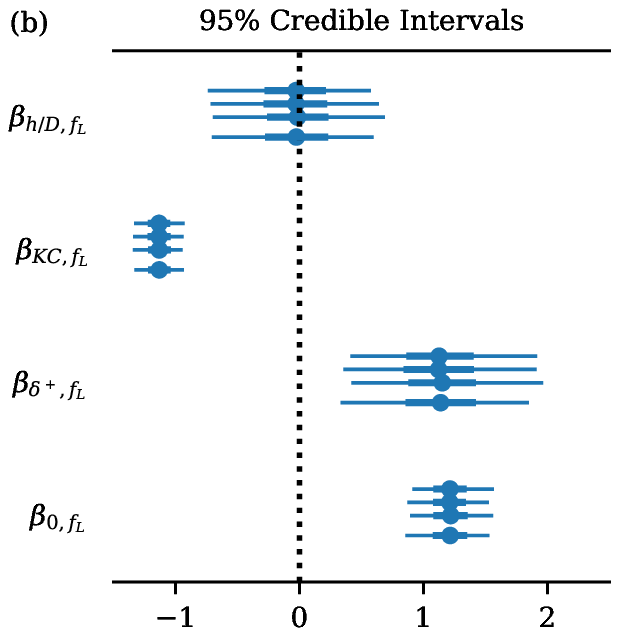}
	}% Images in 100% size
	\caption{Forest plot showing mean value (dot), inter-quartile range (thick line) and 95\% confidence interval of the posterior probability distributions for the scaling parameters of (a) $f_U$ and (b) $f_L$}
	\label{fig:ForestFull}
\end{figure}

By assuming flat (uninformed) prior distributions and normally distributed errors in the MCMC method, the MCMC and ordinary least squares methods are functionally equivalent and yield similar results (Table \ref{tab:MLE}). An informative graphical representation of the MCMC coefficient estimates and relative size of confidence intervals is presented in a Forest plot (figure \ref{fig:ForestFull}) which shows the mean, inter-quartile range and 95\% confidence intervals of the posterior probability distributions. This highlights small effect sizes for $h/D$ and $KC$ in $f_U$ and $h/D$ in $f_L$. This model has a coefficient of determination of $r^2=0.88$ for both $\langle \overline{v} \rangle _{rms}/U_0$ and $\langle \overline{w} \rangle _{rms}/U_0$, however with large confidence intervals for $h/D$, $\delta^+$ and the intercepts.
%Substituting the mean MCMC parameter estimates into Equations \ref{eqn:fu} and \ref{eqn:wU0} gives scaling relationships with high coefficients of determination $r^2=0.88$ for both $\langle v \rangle_{rms}/U_0$ and $\langle w \rangle_{rms}/U_0$ (figure \ref{fig:ScalingFull}). 

% \begin{figure}
% 	\centerline{
% 		\includegraphics[width=1.0\textwidth]{Chapter_4/Figs/HMC_Scaling_Full.eps}
% 	}% Images in 100% size
% 	\caption{Scaling relationships for (a) $\langle \overline{v} \rangle_{rms}/U_0$ and (b) $\langle \overline{w} \rangle_{rms}/U_0$ based on $U_e/U_0\sim f_U(h/D,KC,\delta^+)$ and $L_e/D\sim f_L(h/D,KC,\delta^+)$. The black line indicates a 1:1 slope.}
% 	\label{fig:ScalingFull}
% \end{figure}

\begin{figure}
	\centerline{
		\includegraphics[width=0.5\textwidth]{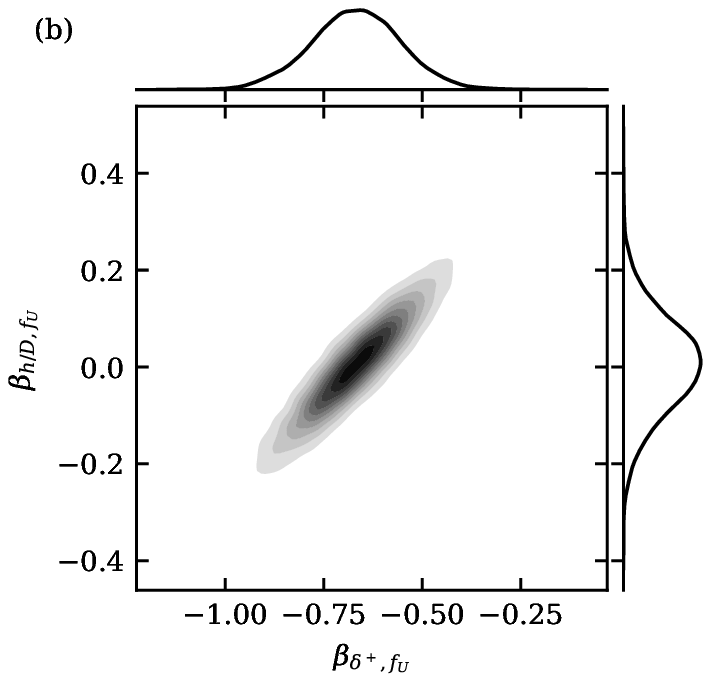}
		\includegraphics[width=0.5\textwidth]{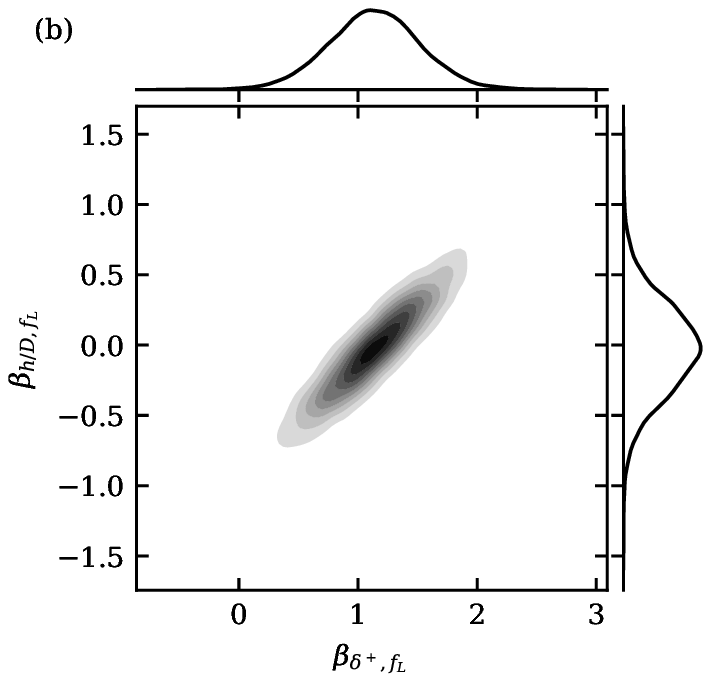}
	}% Images in 100% size
	\caption{Joint posterior probability distributions between scaling parameter estimates for $h/D$ and $\delta^+$ in (a) $f_U$ and (b) $f_L$}
	\label{fig:JointFull}
\end{figure}

\begin{figure}
	\centerline{
		\includegraphics[width=0.5\textwidth]{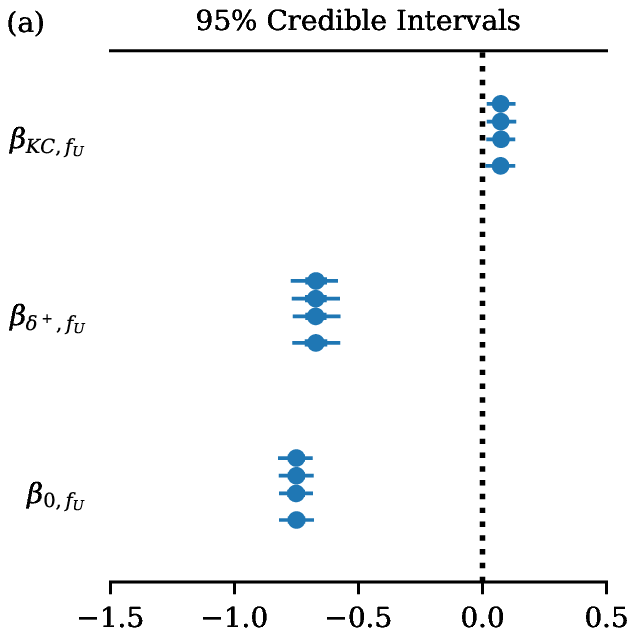}
		\includegraphics[width=0.5\textwidth]{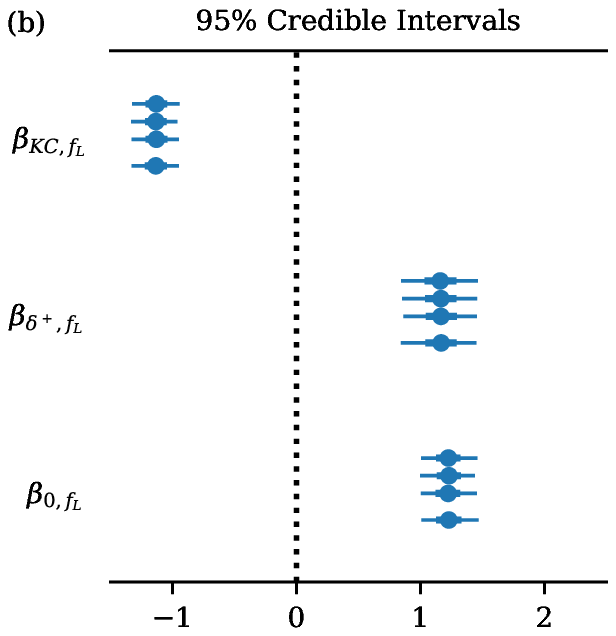}
	}% Images in 100% size
	\caption{Forest plot showing mean value (dot), inter-quartile range (thick line) and 95\% confidence interval of the posterior probability distributions for the scaling parameters of (a) $f_U(KC,\delta^+)$ and (b) $f_L(KC,\delta^+)$}
	\label{fig:ForestRed}
\end{figure}

\begin{figure}
	\centerline{
		\includegraphics[width=0.95\textwidth]{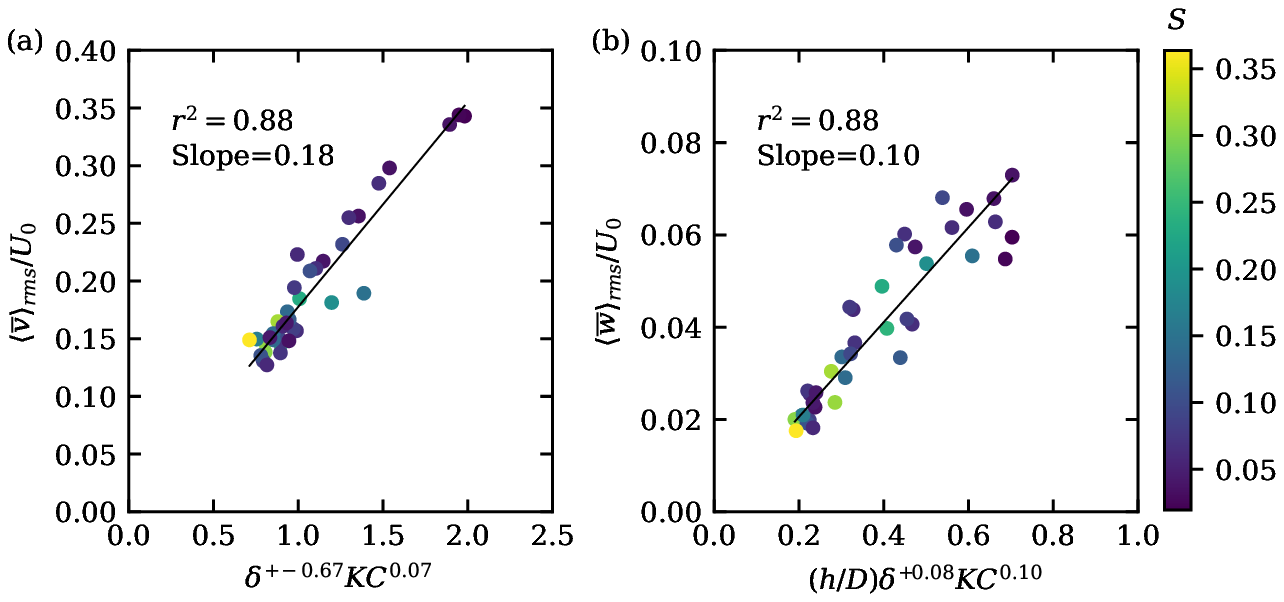}
	}% Images in 100% size
	\caption{Scaling relationships for (a) $\langle \overline{v} \rangle_{rms}/U_0$ and (b) $\langle \overline{w} \rangle_{rms}/U_0$ based on the reduced parametrisation of $U_e/U_0\sim f_U(KC,\delta^+)$ and $L_e/D\sim f_L(KC,\delta^+)$. The black line indicates a 1:1 slope.}
	\label{fig:ScalingRed}
\end{figure}

An advantage of the Bayesian statistical approach implemented with the MCMC method is the ability to examine the joint probability of the posterior distributions of the scaling coefficients. The joint-marginal distributions provide insight into the structure and potential correlation between scaling coefficient estimates. In addition to the small effect size of $h/D$ (figure \ref{fig:ForestFull}), there is significant correlation between the scaling coefficient for $h/D$ and $\delta^+$ in $f_U$ and $f_L$ (figure \ref{fig:JointFull}). This is not surprising given that the flow depth $h$ is present in both $\delta^+$ and $h/D$. Due to the low effect size of $h/D$ and correlation with $\delta^+$, it is justified to remove dependence on $h/D$ from the models of $f_U$ and $f_L$. Removing dependence on $h/D$ significantly narrows the confidence intervals for the remaining coefficients (figure \ref{fig:ForestRed}) without reducing the predictive capacity of the model, $r^2=0.88$ (figure \ref{fig:ScalingRed}). 

The influence of $KC$ on $\langle v \rangle_{rms}/U_0$ and of $\delta^+$ and $KC$ on $\langle w \rangle_{rms}/U_0$ remains weak and to first-order $\langle v \rangle_{rms}/U_0 \sim \delta^{+\!-2/3}$ and $\langle w \rangle_{rms}/U_0 \sim h/D$ (as would be predicted based on conservation of mass arguments). The tidally-forced, shallow island wake has weak non-linearity (with respect to upwelling and lateral velocity fluctuations). However, there is sufficient non-linearity to establish a range of wake forms (symmetric through vortex shedding) with a variety of complex flow structures that influences the spatial distribution of upwelling significantly.

